\documentclass[aip,jcp,reprint]{revtex4-1}
\usepackage{mathtools}
\usepackage{amsmath}
\usepackage{multirow}
\usepackage[english]{babel}
\usepackage{amsmath,lipsum}
\usepackage{etoolbox}
\newcommand{\zerodisplayskips}{%
  \setlength{\abovedisplayskip}{5pt}%
  \setlength{\belowdisplayskip}{5pt}%
  \setlength{\abovedisplayshortskip}{5pt}%
  \setlength{\belowdisplayshortskip}{5pt}}
\appto{\normalsize}{\zerodisplayskips}
\appto{\small}{\zerodisplayskips}
\appto{\footnotesize}{\zerodisplayskips}
\usepackage{graphicx}
\usepackage{color}
\usepackage{xcolor}
\usepackage{ulem}
\usepackage{caption}
\usepackage{subcaption}

\makeatletter
\newcommand*{\addFileDependency}[1]{
  \typeout{(#1)}
  \@addtofilelist{#1}
  \IfFileExists{#1}{}{\typeout{No file #1.}}
}
\makeatother

\renewcommand\thesubsection{\Alph{subsection}}
\renewcommand\thesubsubsection{\arabic{subsubsection}}
\makeatletter
    \def\@seccntformat#1{\@ifundefined{#1@cntformat}%
      {\csname the#1\endcsname\space}
      {\csname #1@cntformat\endcsname}}
    \def\subsection@cntformat{\thesection.\thesubsection\space} 
    \def\subsubsection@cntformat{\thesection.\thesubsection.\thesubsubsection\space}
\makeatother

\DeclareUnicodeCharacter{0301}{\'{e}} 

\begin{document}

\preprint{AIP/123-QED}

\title{Quantifying Energetic and Entropic Pathways in Molecular Systems}
\author{E. R. Beyerle}
 \affiliation{Institute for Physical Science and Technology, University of Maryland, College Park, Maryland 20740, USA}
\author{Shams Mehdi}
\affiliation{Biophysics Program and Institute for Physical Science and Technology, University of Maryland, College Park 20742, USA}
\author{Pratyush Tiwary}
 \email{ptiwary@umd.edu}
 \affiliation{Department of Chemistry and Biochemistry and Institute for Physical Science and Technology, University of Maryland, College Park 20742, USA}

\begin{abstract}
When examining dynamics occurring at non-zero temperatures, both energy and entropy must be taken into account while describing activated barrier crossing events. Furthermore, good reaction coordinates need to be constructed to describe different metastable states and the transition mechanisms between them. Here we use a physics-based machine learning method called the State Predictive Information Bottleneck (SPIB) to find non-linear reaction coordinates for three systems of varying complexity. The SPIB is able to predict correctly an entropic bottleneck for an analytical flat-energy double-well system and identify the entropy- and energy-dominated pathways for an analytical four-well system. Finally, for a simulation of benzoic acid permeation through a lipid bilayer, SPIB is able to discover the the entropic and energetic barriers to the permeation process. Given these results, we thus establish that SPIB is a reasonable and robust method for finding the important entropy and energy/enthalpy barriers in physical systems, which can then be used for enhanced understanding and sampling of different activated mechanisms.  
\end{abstract}
\date{\today}

\maketitle

\section{Introduction}
\label{intro}
The separation of driving forces for a generic chemical mechanism into its energetic and entropic components has been a topic of continued interest over the decades. This is relevant to diverse problems such as solute aggregation,\cite{Mondal2011, Choudhury2006}; drug, ligand binding to proteins \cite{Freire2008, Ladbury2010} and other biomolecules\cite{JenJacobson2000, Starikov2012}; nucleation\cite{Black2007, Radhakrishnan2003}; molecular permeation through membranes \cite{Jo2010, Marrink1994, Maccallum2006}; and entropy-driven phase transitions \cite{Frenkel1993, Frenkel2015, Lee2019, Vo2022}.
 This separation gives fundamental insight into the nature of interactions stabilizing or destabilizing a given material, and can guide further design strategies. 
 
For systems evolving at a non-zero temperature, the energy of the system alone does not determine the stability of different configurations. In addition to the change in energy, $\Delta U$, or enthalpy, $\Delta H$, depending on whether the system is at constant volume or constant pressure conditions, an entropic contribution, $-T\Delta S$, driven by the system's temperature, must also be taken into account. Restricting our attention for the sake of argument to systems at constant pressure that adhere to Kramers' rate law, the reaction rate at a temperature $T$ is proportional to the exponential of the free energy barrier $\Delta G$:\cite{Hanggi1990} $k\propto \exp\left[-\frac{\Delta G}{k_BT}\right] = \exp\left[\frac{-\Delta H}{k_BT}\right]\exp\left[\frac{\Delta S}{k_B}\right] $ where $k_B$ is Boltzmann's constant. One can then disentangle entropic and enthalpic contributions by noting that the rate thus defined can be viewed as the product of a temperature independent term driven by entropy, with all enthalpic contributions restricted to the temperature-dependent term:
\begin{equation}
\ln\left(\frac{k}{k^{\prime}}\right) = \frac{\Delta H}{k_B}\left(\frac{1}{T^{\prime}} - \frac{1}{T}\right),
\label{arrhenius2}
\end{equation}
where $k$, $k'$ denote rates at two temperatures $T$ and $T'$.

However, using Eq. \ref{arrhenius2} to calculate the enthalpic and entropic contributions to the free energy is burdensome, as it requires running experiments or simulations at multiple temperatures, which is likely difficult for myriad reasons. Furthermore, this approach would still not give insight into the molecular origins of the enthalpic and entropic parts of the rate constant. Thus, there is clear need to develop theoretical and computational frameworks that can give direct molecular and atomic level understanding of entropic and enthalpic or energetic driving forces without having to repeat expensive all-atom simulations at different temperatures.

In this work we demonstrate a computational framework that allows quantifying the energetic and entropic contributions to a given chemical process of interest from all-atom unbiased or biased molecular dynamics (MD) simulations performed at a single temperature. Given the importance of this problem,\cite{Guarnera2016, Maccallum2006, Jo2010, Gimondi2018, tsai2019reaction, Katiyar2021} several other approaches have been proposed; however, we would argue that no single approach addresses under the same umbrella both of the following key challenges:
\begin{enumerate}
    \item Chemical processes often involve a multiplicity of pathways \cite{Bryngelson1995, Onuchic1997, Zhuravlev2010, Lee2019, Leoni2021, Jiang2018, Jiang2019, Schwantes2013, Beauchamp2012,Guarnera2016, tsai2021sgoop, Finney2021} with differing energetic and entropic contributions. The approach should be able to quantify these for the different pathways separately instead of just one overall trend and learn low-dimensional  descriptors corresponding to them.
    \item Many processes are effectively rare events when simulated in all-atom femtosecond resolution. Simulating these thus require specialized enhanced sampling methods, which work better if one has prior knowledge of approximate reaction coordinates for the different slow processes mentioned in the challenge above\cite{Tiwary2016,tiwary2013accelerated,Bussi2020}. 
\end{enumerate}

Commonly used dimensionality reduction techniques such as time-lagged independent component analysis (TICA)\cite{Hyvarinen2001}, or Markov state models \cite{Bowman2013} can assist with the first challenge above as they can ascertain the dominant slow modes in a given system. However, they need prior access to extensive sampling, thereby not offering a solution to the second challenge. We seek an approach that is able to learn such slow, multi-dimensional degrees of freedom corresponding to different pathways from preliminary biased or unbiased data, and perform further biased sampling to enhance fluctuations along these pathways \cite{Valsson2016}.

Here, we utilize the state predictive information bottleneck (SPIB) \cite{Wang2021a} that meets both of these challenges together. SPIB has the ability to both find relevant multi-dimensional reaction coordinates and perform enhanced sampling along them, even for simulations of rare events.  We show here how it can be used to learn reaction coordinates for a set of systems that are known to possess distinct entropic and energetic barriers, clearly demarcating different pathways and their respective energetic/entropic components.

The SPIB protocol finds the relevant reaction coordinates by passing the input order parameters through a modified variational autoencoder \cite{Alemi2017} and enhances the sampling of the barrier regions by running metadynamics \cite{Laio2002, Barducci2007} along the optimized SPIB latent coordinate(s). This approach is in contrast to pure TICA, which does not intrinsically enhance the sampling along the discovered slow reaction coordinate(s), and umbrella sampling \cite{Torrie1977}, which enhances the sampling without discovering the reaction coordinate. Furthermore, SPIB, like its predecessor RAVE (re-weighted autoencoded variational Bayes) \cite{Ribeiro2018, Pant2020}, is used to iteratively update the discovered reaction coordinate to enable the sampling of rare events with a feasible amount of compute time. That is, rounds of SPIB can be performed to iteratively optimize the discovered reaction coordinates\cite{Mehdi2021}.

We show that without much prior knowledge the SPIB approach is able to separate different pathways and distinguish the primarily entropic from the primarily energetic pathway. We demonstrate this result for a pair of analytical potentials as well as in the description of benzoic acid permeation through a membrane. We compare the SPIB results to TICA on all problems clearly demonstrating the advantage of using SPIB for the systems studied here. These results indicate the SPIB approach is useful for finding the entropic and energetic reaction coordinates, even though there is no explicit accounting of either the energy or entropy in the SPIB approach. Combined with the SPIB's ability to discover and enhance the sampling along these entropic and energetic reaction coordinates when coupled with metadynamics, we propose the SPIB as a powerful protocol to sample free energy barriers, no matter their thermodynamic origins.


\section{Methods}
\subsection{State Predictive Information Bottleneck (SPIB)}
\label{spib}
The formalism for the SPIB was laid out in Ref. \onlinecite{Wang2021a}, and the method is an extension of the previously developed reweighted autoencoded variational Bayes (RAVE) technique \cite{Ribeiro2018, Wang2020}. Briefly, the SPIB takes as input a set of coordinates from a time-ordered, dynamical trajectory $\mathbf{X}(t)$, and finds a reduced representation of the dynamics $\mathbf{z}(t)$ that maximizes the following loss function $\mathcal{L}$, which can be seen as related to the information bottleneck loss function $\mathcal{L}_{\text{IB}}$ and hence the difference of two mutual information terms as follows\cite{Wang2021a}:
\begin{align}
 \label{loss}
\mathcal{L}_{\text{IB}} & \equiv    I(\textbf{z},\textbf{y}) -\beta I(\textbf{X},\textbf{z}) \\ \nonumber
 & \ge \sum_{k = 1}^{N}\log\left(p(\mathbf{y}({k + s}) | \mathbf{z}(k))\right) - \beta\log\left(\frac{p(\mathbf{z}(k) | \mathbf{X}(k))}{p(\mathbf{z}_{\theta})}\right).
\end{align}

Maximization of the loss function in Eq. \ref{loss} ensures that the SPIB discovers a low-dimensional, compressed representation of the input coordinates that is maximally predictive of the \emph{state} of the system, $\mathbf{y}$, at a lagtime $s$ in the future, $\mathbf{y}({k + s})$. The parameter $\beta \in \left[ 0, \infty\right)$ serves the same function as in a traditional variational autoencoder \cite{Kingma2014, Alemi2017}; tuning $\beta$ governs the trade-off between how compressed the latent representation $\mathbf{z}$ is and how faithfully the latent representation is able to predict the future state of the system $\mathbf{y}({k + s})$. That is, the second term in Eq. \ref{loss} effectively serves as a regularization term \cite{Goodfellow2016} that penalizes a high-dimensional latent space.


The SPIB encoder, $p(\mathbf{z}(k) | \mathbf{X}(k))$, and decoder, $p(\mathbf{y}({k + s}) | \mathbf{z}({k}))$, are generated by a fully-connected, nonlinear neural network, as described in Ref. \onlinecite{Wang2021a}, with two encoding layers and two decoding layers, in addition to the bottleneck layer. The parameter-informed prior $p(\mathbf{z}_{\theta})$ is generated using a variation of the VampPrior proposed in \onlinecite{Tomczak2018}:
\begin{equation}
p(\mathbf{z}_{\theta}) = \sum_{i=1}^K w_i p_{\theta}\left(\mathbf{z} | \mathbf{u}_i\right),
    \label{vampprior}
\end{equation}
where $K$ is the predicted number of states in the system, $w_i$ are the weights of the representative inputs $\mathbf{u}_i$, each of which is, in practice, a single sample of $\mathbf{X}$ selected from each of the $K$ states $\mathbf{y}$. The prior given in Eq. \ref{vampprior} is updated following each refinement of the network as the predicted state labels $\mathbf{y}$ are refined and updated. Furthermore, as the SPIB is iterated, some of the initial state labels $\mathbf{y}(0)$ may merge, depending on the lagtime $s$ selected for the SPIB analysis. That is, as the lagtime $s$ of the SPIB is increased, the anticipated number of states $K$ will decrease. This is because faster motions, corresponding to small barrier crossings are coarse-grained out due to the longer lagtimes. This effect of finding only the most metastable states of the system, corresponding to the highest barrier crossing events, is analogous to what occurs in Markov state modelling of dynamical systems \cite{Bowman2013} when a spectral clustering algorithm such as robust Perron cluster-cluster analysis \cite{Deuflhard2005} is applied to the obtained eigenvectors of the transition operator on the state space. Once the state labels $\mathbf{y}$ converge, the training process is terminated, and the obtained $\mathbf{z}$ is analyzed.

\subsection{Calculating entropy and other state functions from SPIB}
\label{entropy}
For analytical potentials or systems in vacuum, the energy $U(t) = U(\mathbf{X}(t))$ of the system at any time $t$ in the simulation can be directly calculated using the system's state $\mathbf{X}(t)$ at time $t$.  For solvated systems the situation is a bit more complicated, as both the solute intramolecular and solute-solvent intermolecular potentials must be taken into account. Following the procedure described in Ref. \onlinecite{Kollias2020}, the change in energy from a reference state, $\Delta U$, is given by
\begin{equation}
\Delta U = \Delta U_{\text{intra}} + \Delta U_{\text{inter, SR}}
\label{dU}
\end{equation}
where $\Delta U_{\text{intra}}$ and $\Delta U_{\text{inter, SR}}$ denote the energy changes due to solute-solute intramolecular and short-range solute-solvent intermolecular interactions respectively. 
In typical MD simulations, $\Delta U_{\text{inter, SR}}$ comes from Lennard-Jones and electrostatic interactions. Furthermore, for systems simulated in the constant number, pressure and temperature (NPT) ensemble, the change in internal energy due to P-V  work performed by the barostat must be included. Here the solute enthalpy $\Delta H$ is calculated instead of $\Delta U$ as follows:
\begin{equation}
\Delta H = \Delta U_{\text{intra}} + \Delta U_{\text{inter, SR}} + P\Delta V,
    \label{enthalpy_bz}
\end{equation}
where $P\Delta V$ is the change in energy due to the work performed by the barostat, where the reference volume used to calculate $\Delta V$ is the volume of the box at the beginning of the simulation.

In systems of practical interest with high free energy barriers, the sampling in the barrier regions can be noisy. Here, instead of using a regular histogram to calculate the free energy and enthalpy along the reaction coordinate, a kernel density estimate (KDE) \cite{Silverman1986} of the probability distribution along each reaction coordinate is utilized, with the choice of Gaussian kernel. In KDE with a Guassian kernel, the probability distribution along an RC is estimated at each value $z$ of the RC using a sum of Gaussian basis functions\cite{Silverman1986}:
\begin{equation}
p(z) = \frac{1}{N}\sum\limits_{k=1}\limits^{N}\frac{1}{\sqrt{2\pi h^2}}\exp\left[-\left(z - z(k)\right)^2/2h^2\right],
\label{kde}
\end{equation}

where $h$ is the selected bandwidth and $N$ is the number of frames in the trajectory. Choosing KDE over regular histogramming effectively amounts to binning using a Gaussian basis set in place of binning with a basis set of indicator functions; using the Gaussian basis set allows for a smoother estimate of the probability density in regions of the free energy surface where the sampling is noisy, which is the case for systems describing rare events where the transition region is sampled infrequently.


To decompose the contribution of the entropic and energetic components along a generic RC $z$, we define first the general definition of geometric free energy along a reaction coordinate, $G(z)$\cite{Hartmann2011}:
\begin{equation}
G(z)=-k_{B} T \ln \left(\int_{R^{n}} e^{ -U(\mathbf{x}) / k_{B} T}\right. \\
\left.\times \delta(\Phi(\mathbf{x})-z) \operatorname{det}(\tilde{G})^{\frac{1}{2}} d \mathbf{x}\right)
    \label{geometric_energy}
\end{equation}

where $\Phi(\mathbf{x})$ is the desired level set of the reaction coordinate where we desire the calculation of the geometric free energy, $\delta(x)$ is the Dirac delta function, and $\tilde{G}$ is the Gram matrix of the transformation $\Phi(\mathbf{x}): \mathbf{x} \rightarrow \mathbf{z}$. The Gram matrix is decomposable into the product of the Jacobian matrix, $\nabla\mathbf{z}$, and its transpose \cite{Lelievre2010}:
\begin{equation}
\tilde{G}=\left(\nabla\mathbf{z}^T\right)\nabla\mathbf{z}
\label{gram}
\end{equation}

where the Jacobian matrix describes how the $n$-dimensional input space $\mathbf{x}$ is stretched or squeezed as it is transformed nonlinearly via the SPIB-defined neural network to generate the $m$-dimensional reaction coordinate $\mathbf{z}$.

The energy along each level set of $\mathbf{z}$, $U(\mathbf{z})$, can be calculated using the formula for averaging over each level set of the reaction coordinate \cite{Hartmann2011}:
\begin{align}
\langle U(z)\rangle_{\Sigma(z)} := U(z) = \frac{1}{N_z}\int &U(\mathbf{x})e^{-U(\mathbf{x})/k_BT} \notag \\ 
&\times\delta\left(\Phi(\mathbf{x}) - z\right)\det\left(\tilde{G}\right)^{\frac{1}{2}}d\mathbf{x} 
\label{Uz}
\end{align}

where $\Sigma(z)$ is the submanifold of the input coordinates constrained to the given value of the reaction coordinate $z$\cite{Hartmann2011}. That is, $\Sigma(z)$ contains the set of all high-dimensional input coordinates in the trajectory mapped to $z \in (z - \epsilon, z + \epsilon)$ by the SPIB encoder, for some small-enough $\epsilon$ $>$ 0. Finally, $N_z$ is a normalization constant that is equal to the number of frames in the trajectory that are mapped to the submanifold $\Sigma(z)$, augmented by the Jacobian:
\begin{equation}
N_{z}=\sum_{k=1}^{N} I_{z}(X(k)) \operatorname{det}(\tilde{G})^{\frac{1}{2}}
\label{normalization_constant}
\end{equation}
where $I_{z}(X(k))$ is an indicator function over $\Sigma_{z}$ which is equal to 1 if $X(k)$ maps to $\Sigma_{z}$ and equal to 0 otherwise.
Finally, the entropy along $z$ is calculated using the thermodynamic identity $\Delta G(z) = \Delta U(z) - T\Delta S(z) \Rightarrow \Delta S(z) = \frac{1}{T}\left(\Delta U(z) - \Delta G(z)\right)$, where $\Delta U(z) = U(z) - U(z_{\text{ref}})$ and $\Delta G(z) = G(z) - G(z_{\text{ref}})$; the reference value $z_{\text{ref}}$, is defined as the value of $z$ that minimizes $U(z)$: $z_{\text{ref}} = \arg\min\limits_{z}U(z)$. Since $-T\Delta S(z)$ gives the entropic contribution to the free energy barrier along $z$, we will plot $-T\Delta S(z)$ when evaluating the entropic contribution to the free energy barrier along a given SPIB reaction coordinate.

\subsection{Analytical potentials}
\label{anly}
\begin{figure}[htb] 
\center
\includegraphics[width=0.7\columnwidth]{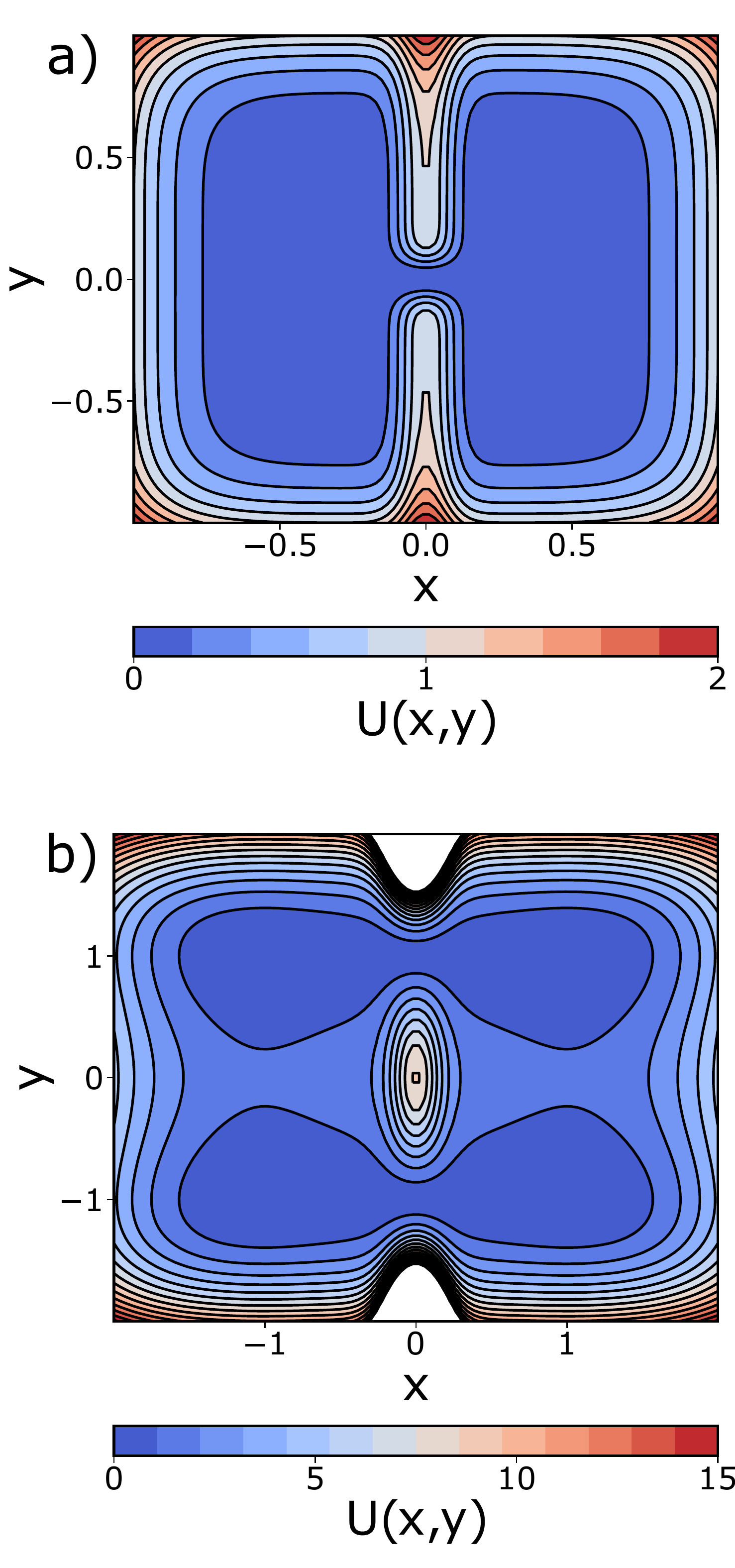}
\caption{Panel a: Potential energy surface for the entropic double-well potential (Eq. \ref{entropic_two_well}). The entropic bottleneck between the two wells is centered at (x, y) = (0.0, 0.0). Panel b: Potential energy surface for the temperature-switch potential (Eq. \ref{t_switch}). Crossing the two channels between the left- and right-hand sides of the potential corresponds to surmounting a barrier that is primarily entropic while crossing the two (broader) channels between the top and bottom wells corresponds to surmounting a barrier that is primarily energetic. The `masked' (white) regions of the landscape correspond to regions of the potential energy surface where U(x,y) $>$ 15.}
\label{analytical_pes}
\end{figure}

We start by considering two analytical potentials. The first system examined is the entropic double-well system described in Ref. \onlinecite{Faradjian2004}, which possesses only an entropic `bottleneck' between two diffusive wells of equal area. This potential energy surface is shown as a contour plot in Figure \ref{analytical_pes}(a) and is defined below:  
\begin{equation}
U(x, y) = x^6 + y^6 + \exp\left[-y^2 / \sigma_y^2\right]\left(1 - \exp\left[x^2 / \sigma_x^2 \right]\right),
    \label{entropic_two_well}
\end{equation}
where $\sigma_x=\sigma_y = 0.1$ define the width of the wells. 

The second analytical model is a four-well system whose slowest dynamics changes from crossing a predominately entropic barrier to a predominately energetic barrier, as described in Ref. \onlinecite{Banisch2020}. This potential energy surface is shown as a contour plot in Figure \ref{analytical_pes}(b) and is defined below:
\begin{equation}
U(x, y) = h_x\left(x^2 - 1\right)^2 + \left(h_y + a((x, \delta)\right)\left(y^2 - 1\right)^2,
    \label{t_switch}
\end{equation}
with $h_x = 0.5$ and $h_y = 1.0$ describing the well width in the x- and y-directions, respectively, $\delta = 0.05$ describing how much the barrier-crossing pathway along the x-direction is squeezed relative to passage along the barrier in the y-direction, and $a(x, \delta)=\frac{1}{5}\left(1-5 \exp \left[-\left(x-x_{0}\right)^{2} / \delta\right]\right)$.

Simulation details regarding the trajectory length in integration timesteps, temperature, and friction coefficient $\gamma$ for the entropic double well and temperature-switch potentials are given in Table \ref{sim_table}. For both analytical potentials, the integration timestep is 0.001 units and the trajectories in each case were saved to file every $\Delta t = 10$ integration steps, giving a total of 1$\times$10$^6$ and 3$\times$10$^6$ frames for analysis, respectively. The SPIB parameters for both the analytical potentials are given in Table SI. Since, \textit{a priori}, the potential is known, the SPIB neural network was trained to find a one-dimensional reaction coordinate for the entropic double well system and a two-dimensional reaction coordinate for the temperature switch system. 

\begin{table}[h!]
{

   \centering
   \caption{Simulation details for the entropic double well (EDW) and temperature-switch (TS) potentials.}
   \label{sim_table}

   \begin{tabular}[c]{l c c}
   Parameter & EDW & TS\\ \toprule

   Integration steps & $1\times 10^7$ & $3\times 10^7$ \\ 
   $\left(\text{k}_B\text(T)\right)^{-1}$ & 10 & 1 \\
   $\gamma$ & 4.0 & 0.5 \\
   \end{tabular}
   }

\end{table}

\subsection{Benzoic Acid Permeation through a DMPC Membrane Bilayer}
\label{BA}
As a more challenging problem, we consider small molecule permeation through membranes, which is an important process for determining the efficacy of pharmaceuticals \cite{Shinoda2016} as well as other biological processes \cite{Lee2016a, Fathizadeh2019}. However, it is known that for a variety of small molecules, membrane permeation cannot be adequately described using a single reaction coordinate \cite{Lee2016a, Fathizadeh2019}. Recently it was shown that using SPIB one can find an adequate reaction coordinate for enhancing the sampling of benzoic acid membrane crossing of a 1,2-dimyristoyl-sn-glycero-3-phosphocholine (DMPC) membrane bilayer\cite{Mehdi2021}.

Building upon the analysis reported in Ref. \onlinecite{Mehdi2021}, we use a set of generic 21 order parameters (OPs) as input to the SPIB analysis. These are the distance vector defined from the centre-of-mass (COM) of aromatic ring in BA to the COM of the lipid bilayer ($\vec{d}_1$), the hydroxl oxygen of the BA to the COM of the bilayer ($\vec{d}_2$), the carbonyl oxygen of BA to the bilayer COM ($\vec{d}_3$), the distance vector from the benzene ring COM to the hydroxyl oxygen ($\vec{d}_4$), and the distance vector defined between the COM of the two leaflets of the bilayer ($\vec{d}_5$). The (x,y,z)-components of $\vec{d}_1$, $\vec{d}_2$, and $\vec{d}_3$ constitute the first 9 OPs. Furthermore, the sines and cosines of the three angles made by $\vec{d}_4$ ($\theta_x,\theta_y,\theta_z$) and $\vec{d}_5$ ($\omega_x,\omega_y,\omega_z$) with the x-, y-, and z-axes of the simulation box are taken as the additional 12 OPs, for a total of 21 OPs. As described in ref. \cite{Mehdi2021} the input to the SPIB analysis is a pair of 25-ns long trajectories where the benzoic acid is initially placed on the positive side of the bilayer membrane in one and on the negative side in the other. A 500-ns, biased simulation along an optimal SPIB one-dimensional reaction coordinate found from this initial pair of 25-ns unbiased molecular dynamics (MD) simulations is analyzed using the SPIB. Full simulation details can be found in the SM or in Ref. \onlinecite{Mehdi2021}. A snapshot from the simulation showing the physical setup of the system, with water molecules excluded, is given in Figure S1 in the SM. 

To perform the enthalpy-entropy decomposition for this system, the enthalpy is calculated using Eq. \ref{enthalpy_bz} with all energies, pressure, and volume calculated using the gmx energy module in GROMACS 2020.2\cite{Abraham2015}. To calculate the free energy, enthalpy, and entropy profiles along each of the two SPIB RCs, the KDE method outlined in Section \ref{entropy} is used with the bandwidths for the first and second RC equaling $h=0.5$ and $h=1.0$ respectively.

\begin{figure}[t] 
\center
\includegraphics[width=1\linewidth]{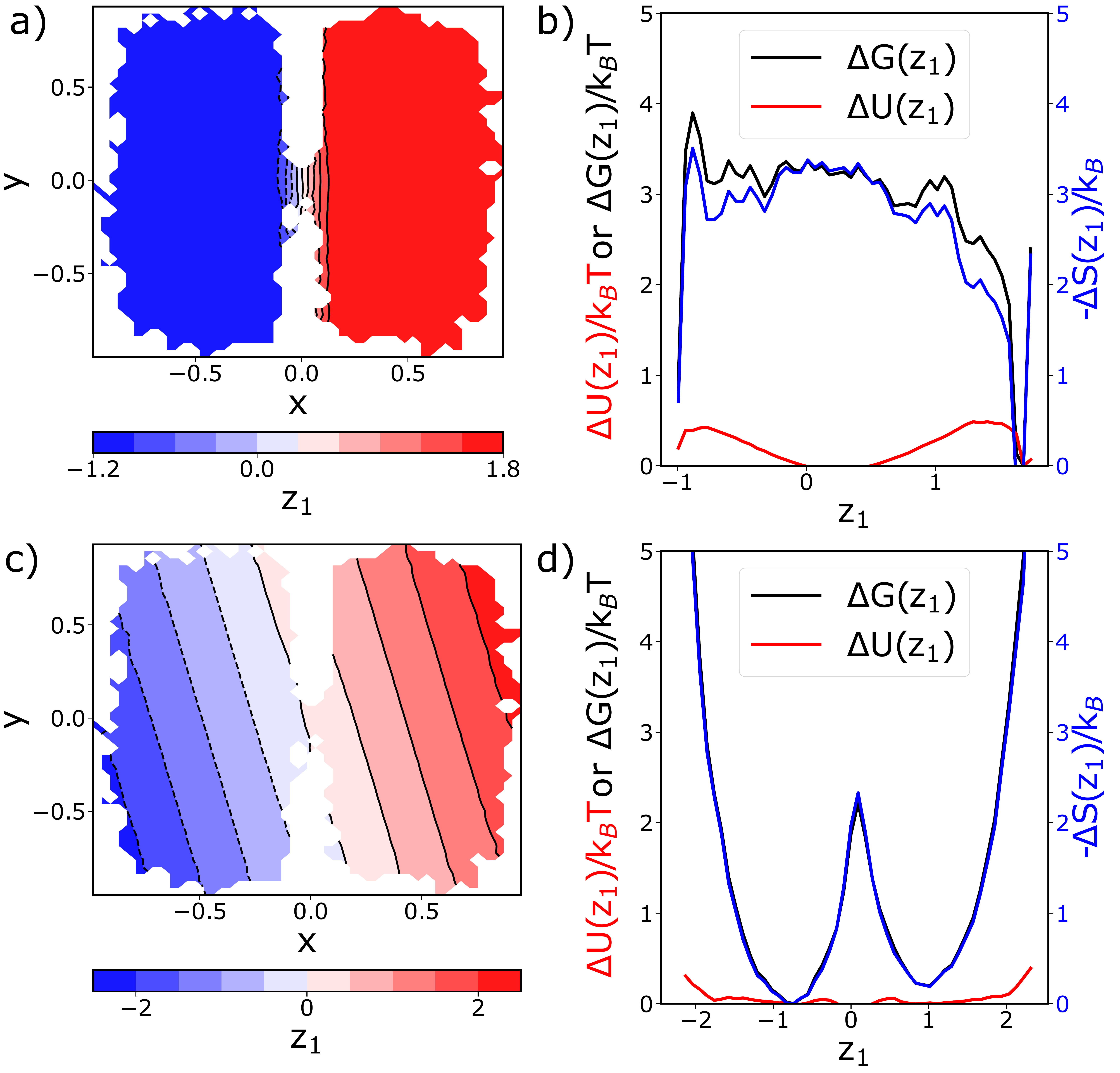}
\caption{Panel a: Projection of the one-dimensional SPIB coordinate, $z_1$, onto the free energy surface for the entropic double-well. Panel b: Plot of $\Delta G(z_1)/k_BT$ (black), $\Delta U(z_1)/k_BT$ (red), and $-\Delta S(z_1)/k_B$ (blue) for the SPIB reaction coordinate. Since the reaction coordinate `expands' the bottleneck region in the latent space, the free energy barrier, which is narrow in (x, y) space, is extended in $z_1$ space. Panel c: The slowest TIC projected onto the entropic double well potential. In contrast to the SPIB, the gradient of the TIC is almost uniform throughout the free energy surface. Panel d: Thermodynamic decomposition of the free energy barrier along the slowest TICA coordinate. Comparison with panel b shows that TICA predicts qualitatively similar results as the SPIB, although the width of the barrier along the slowest TIC is much narrower than in the SPIB latent space, again due to the uniform gradient of the slowest TIC.}
\label{EDW_square}
\end{figure}
\section{Results}
\subsection{Entropic Double Well System}

\subsubsection{SPIB Discovers the Slow, Entropy-dominated Process}
\label{edwA}
The relevant results for the entropic double well system are presented concisely in Figure \ref{EDW_square}, where Figure \ref{EDW_square}(a) shows the one-dimensional SPIB coordinate, $z_1$, for the two-dimensional entropic well system projected on the underlying (x,y) space. SPIB finds a RC that is constant inside the diffusive and energetically constant wells to the left and right of x = 0.0, while $z_1$ changes very rapidly inside the bottleneck between wells. This sign change of $z_1$ inside the bottleneck indicates that $z_1$ describes transitions between the diffusive wells through the bottleneck as the relevant, slow process occurring in the system.  The profiles of $\Delta G(z_1)$, $\Delta U(z_1)$, and $T\Delta S(z_1)$ along the SPIB coordinate $z_1$ for the entropic double-well system are plotted in Figure \ref{EDW_square}(b). It is clear that the transition through the bottleneck is entropy driven. However, since the real-space volume of the wells is large (i.e. a single level set of $z_1$ in Figure \ref{analytical_pes}(a) covers the left- and right-hand wells, respectively), the largest magnitude values of $z_1$ are favorable entropically. We would like to point out that edge effects in $\Delta U(z_1)$ are due to thermal fluctuations pushing the system slightly up the sextic potential when sampling outside the bottleneck. 

\subsubsection{Comparing SPIB with TICA}
\label{edwB}
For comparison to a reference method to extract the slow dynamics of a system, we run TICA\cite{Hyvarinen2001, Perez-Hernandez2013} on the same trajectory fed to the SPIB. The theory and applications of TICA to physical systems has been covered extensively over the past decade\cite{Perez-Hernandez2013, Husic2016, Naritomi2011, Beyerle2021c}.

As with the SPIB analysis of the entropic double-well trajectory, only a single TIC is output from the TICA, which is run using a lagtime of $\tau$ = 10$\Delta t$. The single TIC is projected onto the entropic double-well potential is given in Figure \ref{EDW_square}(c). At first glance, the result is similar to the SPIB result. The slowest TIC predicts transitions between the left and right wells through the bottleneck to be the slowest process in the system. However, the SPIB and TICA results differ in a crucial aspect.  SPIB due to its non-linear nature gives much superior resolution in the entropic bottleneck region. On the other hand, the gradient of the slowest TIC is constant over the entire potential energy surface, as it is a linear method. Thus, the input space is neither stretched or squeezed during TICA. In contrast, SPIB's non-linearity allows for compression inside the bottleneck and expansion in the wells, where the dynamics is fast and mostly orthogonal to the slow, barrier-crossing process. From the thermodynamic perspective, TICA is on par with the SPIB in predicting both that the barrier to transition between the wells is completely dominated by the entropy contribution. This agreement is demonstrated by comparing the results for SPIB in Figure \ref{EDW_square}(d) with the results from TICA in Figure \ref{EDW_square}(b).

\subsubsection{Committors from SPIB and MSM}
\label{edwC}
Finally, to quantitatively assess the quality of the SPIB-predicted reaction coordinate for this system, the committor function \cite{E2006, Noe2009, Bowman2013} is calculated from a Markov state model (MSM) \cite{Noe2007} with a lagtime $\tau_{\text{MSM}}=10\Delta t$ constructed on the (x,y) state space discretized into 500 states using a regular space clustering algorithm \cite{Prinz2011,Scherer2015}. To define the committor function on this discrete space, the `reactant state' or `state A' is defined as the set of all (x,y) mapped to $\widehat{z}_1 = \frac{z_1 - \min(z_1)}{\max(z_1) - \min(z_1)}\le 0.02$ by the SPIB neural network and the `product state' or `state B' as the set of all (x,y) mapped to $\widehat{z}_1 > 0.98$; $z_1$ is transformed to its `min-max' version $\widehat{z}_1$ because, for two-state systems, the committor function should map onto this rescaled version of the coordinate describing the crossing between the wells \cite{Berezhkovskii2004, Buchete2008}.

\begin{figure}[t] 
\center
\includegraphics[width=1\linewidth]{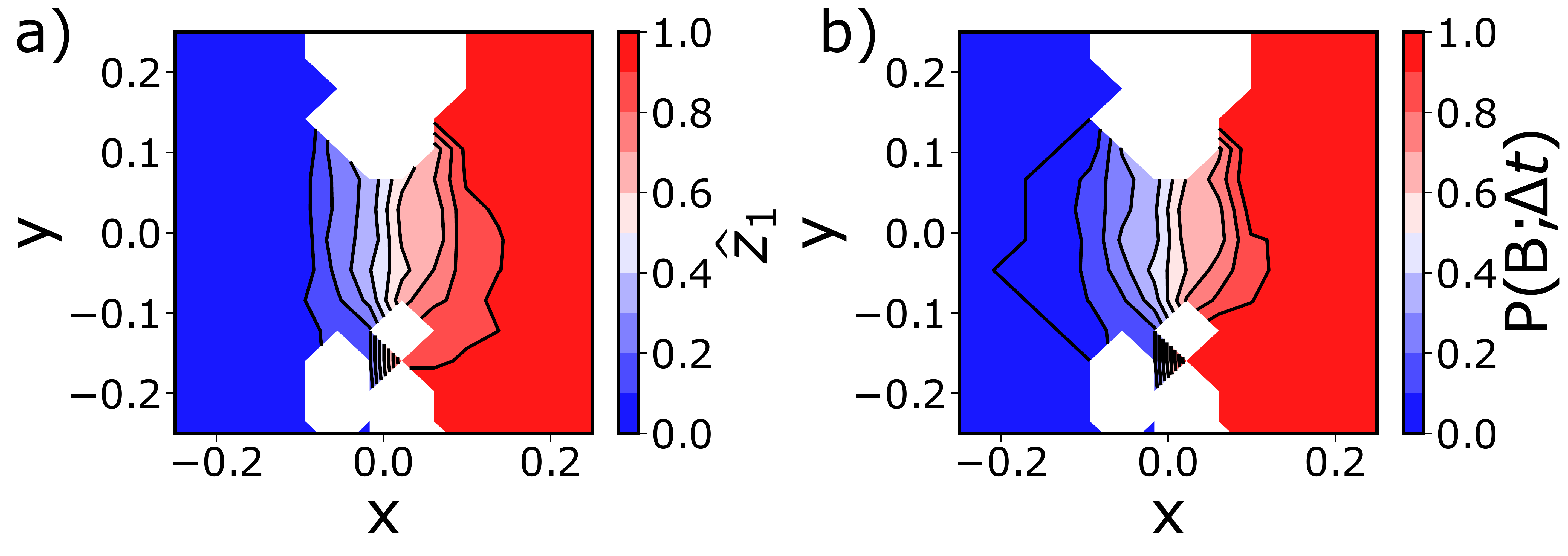}
\caption{Panel a: Projection of $\widehat{z}_1$ in the bottleneck region of the entropic double well potential energy surface. Panel b: Committor function predicted from the MSM with 500 discrete states and lag $\tau_{\text{MSM}} = 10\Delta t$ projected onto the bottleneck region of the entropic double well potential energy surface.}
\label{committor}
\end{figure}

Figure \ref{committor} compares $\widehat{z}_1$ i.e. a linearly scaled version of the SPIB coordinate,   and the committor function calculated using the MSM. Figure \ref{committor}(a),(b) emphasizes the comparison in the bottleneck or transition region of the free energy surface, where both $\widehat{z}_1$ and the committor function change rapidly. With the exception of some `bowing' of the isocommittor surfaces at the entrance to the bottleneck, there is little quantitative difference between $\widehat{z}_1$ and the committor function calculated using the MSM. This result indicates that SPIB is able to accurately and directly learn the committor function for this analytical system with an entropic barrier. 

\subsection{Temperature Switch Potential}

\subsubsection{SPIB Separates the Entropy- and Energy-dominant Pathways}
\label{tsA}
A more challenging, two-dimensional, analytical system is given by the temperature-switch potential presented in Ref. \onlinecite{Banisch2020}. As shown in Ref. \onlinecite{Banisch2020}, an interesting feature of this system is that the slowest process changes from crossing the energetic barrier along the y-component at low temperatures (where the $\Delta U$ contribution to $\Delta A$ dominates) to crossing the entropic barrier at high temperatures (where the $-T\Delta S$ term in $\Delta A$ dominates), hence the moniker `temperature-switch' for this system. Here we are testing if SPIB can select the correct two-dimensional reaction coordinate that captures these two different energetic and entropic pathways at an arbitrary temperature of $\left(k_BT\right)^{-1} = 1.0$.

\begin{figure}[b] 
\center
\includegraphics[width=1\linewidth]{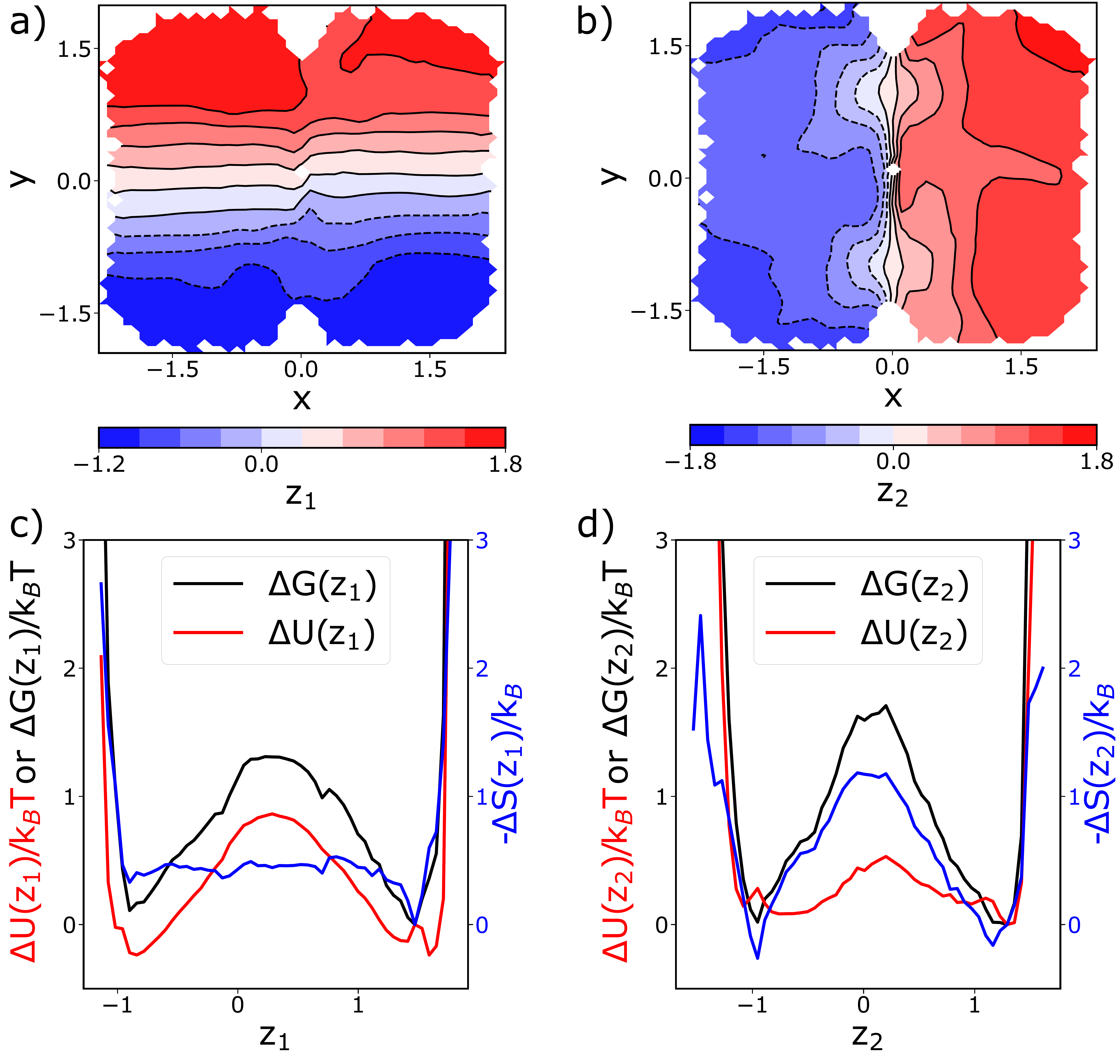}
\caption{Panel a: projection of $z_1$ onto the free energy surface of the temperature switch system. At the given temperature, the first SPIB coordinate describes crossing of the energetic barrier in the y-direction.  Panel b: projection of $z_2$ onto the free energy surface of the temperature switch system. At the given temperature, the second SPIB coordinate describes crossing of the entropic barrier in the x-direction. Panel c: decomposition of the free energy profile (black) along $z_1$ into its energetic (red) and entropic (blue) components. Based on this decomposition, it is clear the barrier along $z_1$ has a majority contribution from the energetic component of the free energy, although there is some `baseline' contribution from the entropy barrier. Panel d: decomposition of the free energy profile (black) along $z_2$ into its energetic (red) and entropic (blue) components. Based on this decomposition, it is clear the barrier along $z_2$ has a majority contribution from the entropic component of the free energy, although there is a small contribution from the energy barrier.}
\label{T-switch_square}
\end{figure}

Panels a and b of Figure \ref{T-switch_square} show both the SPIB latent space coordinates projected individually onto the (x,y) coordinates. Figure \ref{T-switch_square}(a) shows that $z_1$ describes transitions across the system's energetic barrier in the y-direction, while Figure \ref{T-switch_square}(b) shows that $z_2$ describes transitions across the entropic barrier in the x-direction. Here as well as in the previous double well system, SPIB demarcates the entropic bottleneck, while also distinguishing the two possible pathways.  Overall, based on the SPIB decomposition presented in Figure \ref{T-switch_square}, we conclude that the SPIB is able to select two reaction coordinates that separate nearly completely the dynamics corresponding to crossing the energetic and entropic barriers in the system. This statement is further quantified by decomposing the free energy barrier into its energetic and entropic components. Figure \ref{T-switch_square}(c),(d) shows the analogous decomposition for both SPIB coordinates discovered for the temperature-switch potential. Along $z_1$ the change in free energy is almost entirely due to potential energy, while along $z_2$ the change in free energy is almost entirely due to entropy, justifying their labeling as the energetic component and entropic components respectively. In addition, the nonlinear SPIB approach allows for greater resolution of the transition regions, where the slow processes are occurring, and decreased resolution in the free energy wells, where the `uninteresting' fast processes are occurring.

\subsubsection{Comparing SPIB with TICA}
\label{tsB}
As with the entropic double-well potential, TICA is performed on the trajectory for this system (see SM for results). We find that the barrier decomposition from TICA is qualitatively similar to the SPIB results, but that the TICA coordinates possess a linear gradient along the original (x,y) coordinates and thus give a poorer resolution of the barrier regions compared to the SPIB reaction coordinates. Thus, we conclude that for this system as well, it is important to have a non-linear reaction coordinate to describe the slow dynamics in the vicinity of the transition state.


\subsection{Benzoic Acid Membrane Permeation through Phospholipid bilayer}
The third and final system studied here is benzoic acid permeation through a DMPC membrane (BA-DMPC). For this system the reaction coordinate and its energetic/entropic components are not \textit{a priori} clear. The problem is further complicated because small molecule permeation through a lipid bilayer is a slow process, usually requiring enhanced sampling methods to calculate transport properties and permeation rates \cite{Lee2016a, Marrink1994, Cardenas2013}. As such, the permeation of benzoic acid through a DMPC bilayer is an adequate example to test the SPIB's ability to both find and accelerate the sampling along the system's reaction coordinates and separate the entropy and energy barriers to permeation.

\subsubsection{Free energy along intuitive, physical projections}
\label{sec:dmpc_physical}

For the BA-DMPC system, the number of input coordinates to the SPIB is 21 (see SM and Ref. \onlinecite{Mehdi2021}) with a two-dimensional SPIB latent space. For seeding the initial SPIB state labels, regular space clustering as implemented in PyEMMA2 \cite{Scherer2015} in the two-dimensional space spanned by (1) the distance of the center-of-mass of the aromatic ring in benzoic acid to the center-of-mass of the membrane bilayer, $d_{1,z}$, and (2) the angle between the bilayer normal and a vector pointing from the center-of-mass of the aromatic ring of benzoic to the center-of-mass of the hydroxyl oxygen in the carboxylic acid functional group, $\theta_z$. All other parameters for the SPIB analysis are given in Table S1 of the SM.

\begin{figure}[t] 
\center
\includegraphics[width=.9\columnwidth]{ 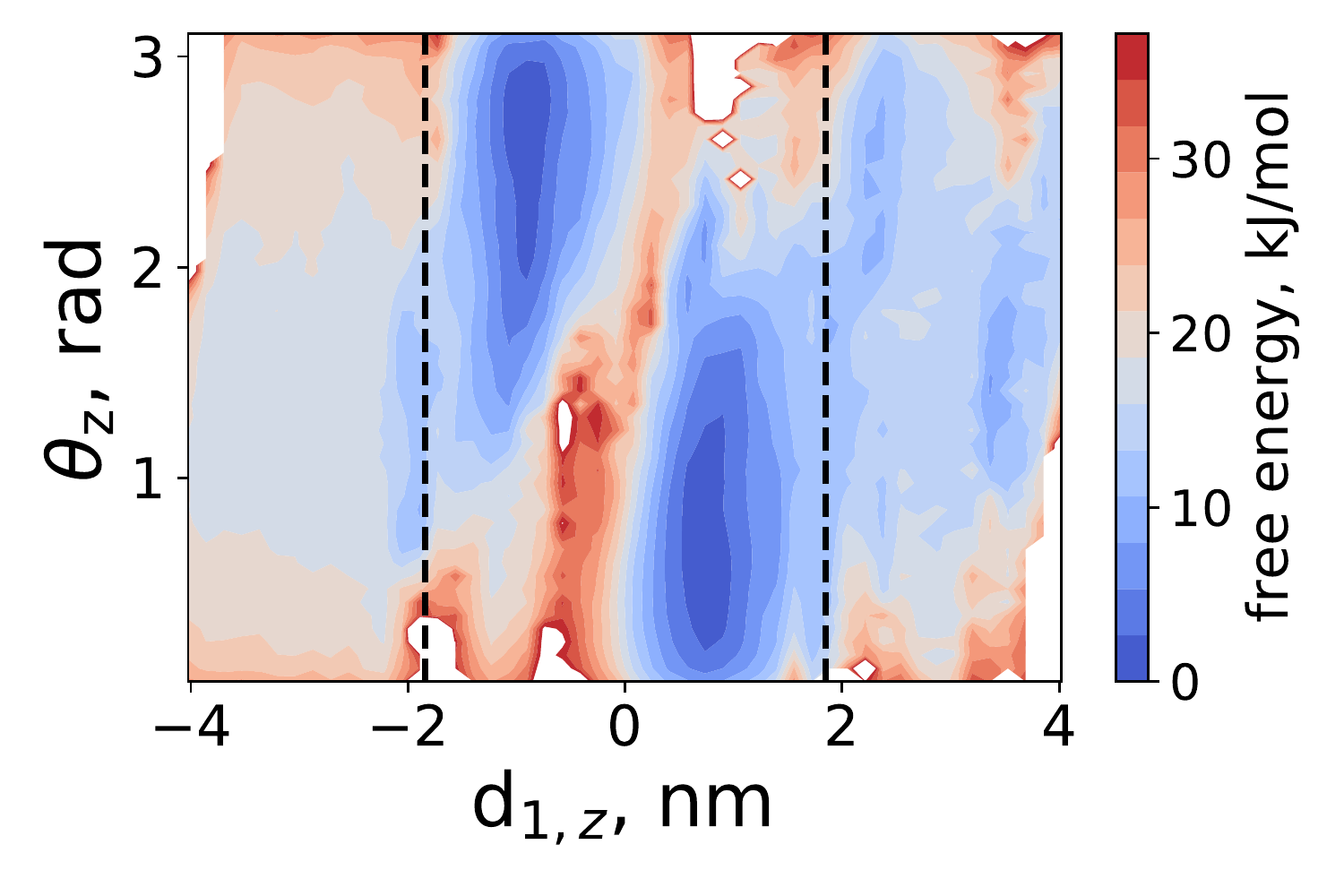}
\caption{Free energy surface spanned by $d_{1,z}$ and $\theta_z$ for the BA-DMPC system. Dashed, vertical, magenta lines are drawn at the average location of the center-of-mass of the phosphorous atoms in the lipid headgroups for the top and bottom layers of the bilayer.}
\label{BA-DMPC_fes}
\end{figure}

Recent research has shown that a one-dimensional projection  along the z-coordinate is not sufficient for describing the permeation process of small molecules through a lipid membrane \cite{Fathizadeh2019}. For example, from the two-dimensional free energy surface given in Figure \ref{BA-DMPC_fes}, it is clear that, to cross inside the membrane, the angular values of $\theta_z$, on average, must be restricted so that the -COOH group in benzoic acid is pointing toward the phospholipid head groups, resulting in an energetically favorable dispersion interaction. However, the restriction of the conformational freedom of the ring after entering the membrane compared with its orientational freedom outside the membrane represents an entropic barrier to membrane crossing. In contrast, the benzoic acid crossing the center-of-mass of the bilayer represents an energetic barrier, since the benzoic acid must lose its favorable dispersion interaction with the headgroup moieties to cross the bilayer. The free energy along these two physical OPs $d_{1,z}$ and $\theta_z$ is given in Figure \ref{BA-DMPC_fes}. This free energy was obtained from a 500-ns biased well-tempered metadynamics simulation along an optimized one-dimensional SPIB reaction coordinate described in Ref. \onlinecite{Mehdi2021}. Since this is a biased trajectory, the contribution from each frame is reweighted appropriately\cite{Barducci2007, Bussi2019} when constructing the free-energy surface in Figure \ref{BA-DMPC_fes}. We emphasize that SPIB is given a larger set of 21 OPs and discovers mechanistically relevant low-dimensional projections on its own.

\subsubsection{SPIB Separates the Entropy and Enthalpy Barriers}

\begin{figure}[htb] 
\center
\includegraphics[width=1\linewidth]{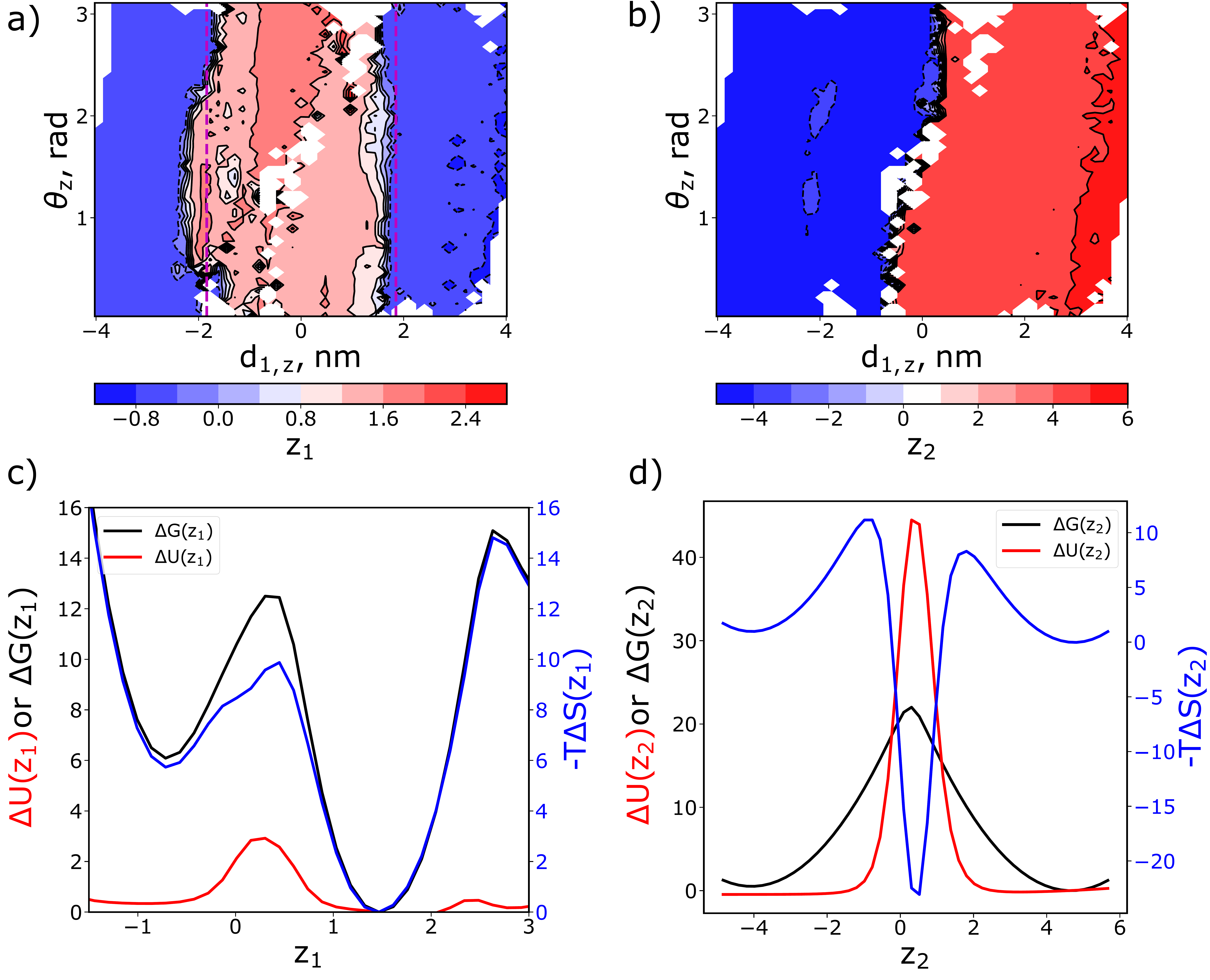}
\caption{Panel a: projection of $z_1$ from the SPIB analysis of the weighted BA-DMPC trajectory onto the surface spanned by the OPs $d_{1,z}$ and $\theta_z$. Panel b: projection of $z_2$ from the SPIB analysis of the weighted BA-DMPC trajectory onto the surface spanned by the OPs $d_{1,z}$ and $\theta_z$. Panel c: projection of the free energy (black), enthalpy (red), and entropy (blue) along $z_1$. The entropy profile nearly traces the free energy profile, indicating that the process described by $z_1$ is entropy dominated. Panel d: same as the left panel, except for $z_2$. Here, the free energy barrier is dwarfed by the enthalpic contribution and there is an entropy-enthalpy compensation effect at the barrier due to the larger accessible volume in the middle of the membrane\cite{Maccallum2006, Marrink1994}. Units of $\Delta G$, $\Delta U$, and $-T\Delta S$ are kJ/mol.}
\label{BA-DMPC_square}
\end{figure}

Figures \ref{BA-DMPC_square}a,b show the two latent space coordinates learned by an SPIB analysis of the biased 500-ns trajectory, projected on the two physical parameters introduced in Sec. \ref{sec:dmpc_physical}. Figures \ref{BA-DMPC_square}c,d show the thermodynamic barrier decomposition along the two SPIB coordinates for the BA-DMPC system. Figure \ref{BA-DMPC_square}(a) shows that the first SPIB coordinate $z_1$ changes sign on both sides of the phospholipid bilayer, indicating this coordinate describes benzoic acid entry into the bilayer. The decomposition of the free energy along $z_1$ into energy and entropy as shown in Figure \ref{BA-DMPC_square}(c) establishes that  $z_1$  describes the entropic process of the permeation mechanism, comprising ligand diffusion into the membrane and subsequent re-orientation. The small energetic barrier along $z_2$ likely corresponds to the unfavorable interactions between the nonpolar aromatic ring of benzoic acid as it passes through the polar phosolipid headgroups and into the interior of the bilayer.

Figure \ref{BA-DMPC_square}(b) shows that the second SPIB coordinate $z_2$ corresponds to the benzoic acid passing through the center of the bilayer. Figure \ref{BA-DMPC_square}(d) establishes that along $z_2$, the free energy barrier is dwarfed by the enthalpic barrier due to an entropy-enthalpy compensation effect when benzoic acid reaches the center of the bilayer. This entropy-enthalpy compensation occurs because there is a larger free volume for the benzoic acid to occupy \cite{Maccallum2006, Marrink1994} in the center of the membrane and no preferential oritentaion of the benzoic acid with respect to the membrane normal due to the loss of the favorable dispersion forces between the benzoic acid and the phospholipid headgroups.  Thus, for this system SPIB is able to separate the majority enthalpic and entropic processes from each other. 

\subsubsection{Comparing SPIB with TICA}
As with the other systems, the SPIB results are compared with the two slowest TICA coordinates for the same trajectory in the SM. We find that the slowest TICA coordinate essentially corresponds to dynamics along the $d_{1,z}$ coordinate and contains both the relevant energetic and entropic barriers in the system. However, most strikingly (Figure S3 in SM), the second TICA coordinate is a fairly non-physical coordinate describing dynamics almost strictly along $\theta_z$, which does not surmount any relevant barriers in the $\left(d_{1,z}, \theta_z\right)$ coordinate space. 

We postulate that this second TICA coordinate is `confused' by some slow, but likely irrelevant process occurring in one or more of the other 18 OPs. This effect is related to the point made previously by others that TICA is susceptible to catching slow, correlated motions that happen to be irrelevant to the interesting dynamics \cite{Sittel2018}. That is, the second TIC must be capturing an irrelevant slow process that has already been projected out in the other 19 OPs and is noise when projected to ($d_{1,z}, \theta_z$) space. 

Overall, we see that TICA is not able to find separate coordinates to describe the entropy and energy barrier crossings in the system, and the second TICA coordinate is not readily interpretable in ($d_{1,z}, \theta_z$). This outcome would have repercussions when using TICA coordinates to perform additional rounds of metadynamics or other biased sampling calculations.

\section{Discussion and Conclusions}
In this work we have shown that the state predictive information bottleneck (SPIB) method is able to extract reaction coordinates (RCs) for systems with energetic, entropic or mixed barriers. The method is demonstrated to work with biased or unbiased simulations and quantifies the precise enthalpic/energetic or entropic contributions to a given activated pathway. Our results show the effectiveness of using a nonlinear method, here SPIB, for finding reaction coordinates RCs as competing linear methods do not often do as satisfactory a job. The separation of reaction coordinates into energetic and entropic components should be important for performing enhanced sampling intelligently for such systems; path-based methods such as forward-flux sampling \cite{Allen2009} or milestoning \cite{Faradjian2004} can be used to push the permeant over the entropic barrier and adaptive biasing methods such as metadynamics \cite{Laio2002} or umbrella sampling \cite{Torrie1977} can be used to push the permeant over the enthalpic barrier at the center of the bilayer.

An additional advantage of using the nonlinear RCs from SPIB is an `adaptive resolution' of the RC, with higher resolution of the transition states and lower resolution of the energetic wells, compared to the linear TICA method; this effect is seen for all three systems examined here (viz. Figures \ref{EDW_square}a,c; \ref{T-switch_square}a,b;   \ref{BA-DMPC_square}a,b; S2a,b; S3a,b). This increased resolution in the transition state is important because it allows a high-fidelity reproduction of the committor function (Figure \ref{committor}) and should allow for better sampling of the transition state during subsequent enhanced sampling simulations biased along the nonlinear RC.

This increased resolution should allow the biased simulation to give better detail regarding the physical and chemical mechanisms occurring at the transition state because its resolution there is finer. That is, the linear method is suitable for finding the qualitatively correct RC, but a nonlinear RC should give better quantitative insight, especially regarding dynamics at the transition state. This effect is similar to what is seen in the Markov state modelling, where increasing the resolution of the indicator function basis set in the transition region greatly reduces the error in the approximation of the slow dynamics and, hence, results in a better model displaying more Markovian dynamics \cite{Prinz2011}.

The BA-DMPC system is a good example of the importance of using a non-linear RC and the SPIB's ability to separate the entropic and enthalpic contributions to a single reaction mechanism, the permeation of a small molecule through a lipid bilayer. When the linear TICA method is used to find the RCs for this system, it finds one useful RC describing the transition from one side of the bilayer to the other, but fails to separate the entropy and enthalpy barriers along this reaction path. Instead, it lumps them into a single, slow RC (Figure S3a). In contrast, the SPIB is able to sift the permeation mechanism into the entropic process (entering and exiting the membrane bilayer) and the enthalpic process (benzoic acid moving from the underside of one leaflet to the underside of the other leaflet, coupled with a reorientation of benzoic acid's -COOH moiety).

Finally, the ability of the SPIB to separate the entropic and enthalpic barriers in a system such as BA-DMPC could still have been partly serendipitous since these two types of barriers happen to separate cleanly for this system. For physical systems where the entropy and energy barriers are more entangled, it cannot be expected that the SPIB will perform so well at distinguishing the thermodynamic origin of the barrier \textit{a priori}. This shortcoming of the method can be circumvented by adding an extra term to the loss function that explicitly forces one RC to traverse a pathway with maximum entropy change and another the pathway with the maximum energy or enthalpy change. This exciting avenue for adding physics-based constraints to training SPIB will be explored in future work.

\section{Acknowledgements}
This research is entirely supported by the U.S. Department of Energy, Office of Science, Basic Energy Sciences, CPIMS Program, under Award DE-SC0021009. Computational resources were provided by Deepthought2, MARCC and XSEDE\cite{Towns2014} (projects CHE180007P and CHE180027P). The authors thank Dedi Wang for useful discussions and feedback regarding the SPIB analysis as well as a critical reading of the manuscript and Sun-Ting Tsai for supplying the codes to perform the Langevin simulations for the entropic double well and temperature switch systems as well as a critical reading of the manuscript. 
\section{Data Availability}
All codes and MD trajectories used to perform the analysis are available following reasonable request to the authors. The code to perform the SPIB analysis is available on GitHub: https://github.com/tiwarylab/State-Predictive-Information-Bottleneck. Codes to calculate the Jacobian and make the plots in the main text for the entropic double well and temperature switch systems can also be found in the following GitHub repository: https://github.com/erb24/jacobian.

\begin{thebibliography}{75}%
\makeatletter
\providecommand \@ifxundefined [1]{%
 \@ifx{#1\undefined}
}%
\providecommand \@ifnum [1]{%
 \ifnum #1\expandafter \@firstoftwo
 \else \expandafter \@secondoftwo
 \fi
}%
\providecommand \@ifx [1]{%
 \ifx #1\expandafter \@firstoftwo
 \else \expandafter \@secondoftwo
 \fi
}%
\providecommand \natexlab [1]{#1}%
\providecommand \enquote  [1]{``#1''}%
\providecommand \bibnamefont  [1]{#1}%
\providecommand \bibfnamefont [1]{#1}%
\providecommand \citenamefont [1]{#1}%
\providecommand \href@noop [0]{\@secondoftwo}%
\providecommand \href [0]{\begingroup \@sanitize@url \@href}%
\providecommand \@href[1]{\@@startlink{#1}\@@href}%
\providecommand \@@href[1]{\endgroup#1\@@endlink}%
\providecommand \@sanitize@url [0]{\catcode `\\12\catcode `\$12\catcode
  `\&12\catcode `\#12\catcode `\^12\catcode `\_12\catcode `\%12\relax}%
\providecommand \@@startlink[1]{}%
\providecommand \@@endlink[0]{}%
\providecommand \url  [0]{\begingroup\@sanitize@url \@url }%
\providecommand \@url [1]{\endgroup\@href {#1}{\urlprefix }}%
\providecommand \urlprefix  [0]{URL }%
\providecommand \Eprint [0]{\href }%
\providecommand \doibase [0]{http://dx.doi.org/}%
\providecommand \selectlanguage [0]{\@gobble}%
\providecommand \bibinfo  [0]{\@secondoftwo}%
\providecommand \bibfield  [0]{\@secondoftwo}%
\providecommand \translation [1]{[#1]}%
\providecommand \BibitemOpen [0]{}%
\providecommand \bibitemStop [0]{}%
\providecommand \bibitemNoStop [0]{.\EOS\space}%
\providecommand \EOS [0]{\spacefactor3000\relax}%
\providecommand \BibitemShut  [1]{\csname bibitem#1\endcsname}%
\let\auto@bib@innerbib\@empty
\bibitem [{\citenamefont {Mondal}\ and\ \citenamefont
  {Yethiraj}(2011)}]{Mondal2011}%
  \BibitemOpen
  \bibfield  {author} {\bibinfo {author} {\bibfnamefont {J.}~\bibnamefont
  {Mondal}}\ and\ \bibinfo {author} {\bibfnamefont {A.}~\bibnamefont
  {Yethiraj}},\ }\bibfield  {title} {\enquote {\bibinfo {title} {{Driving Force
  for the Association of Amphiphilic Molecules}},}\ }\href {\doibase
  10.1021/jz201046x} {\bibfield  {journal} {\bibinfo  {journal} {The Journal of
  Physical Chemistry Letters}\ }\textbf {\bibinfo {volume} {2}},\ \bibinfo
  {pages} {2391--2395} (\bibinfo {year} {2011})}\BibitemShut {NoStop}%
\bibitem [{\citenamefont {Choudhury}\ and\ \citenamefont {{Montgomery
  Pettitt}}(2006)}]{Choudhury2006}%
  \BibitemOpen
  \bibfield  {author} {\bibinfo {author} {\bibfnamefont {N.}~\bibnamefont
  {Choudhury}}\ and\ \bibinfo {author} {\bibfnamefont {B.}~\bibnamefont
  {{Montgomery Pettitt}}},\ }\bibfield  {title} {\enquote {\bibinfo {title}
  {{Enthalpy-entropy contributions to the potential of mean force of nanoscopic
  hydrophobic solutes}},}\ }\href {\doibase 10.1021/jp056909r} {\bibfield
  {journal} {\bibinfo  {journal} {Journal of Physical Chemistry B}\ }\textbf
  {\bibinfo {volume} {110}},\ \bibinfo {pages} {8459--8463} (\bibinfo {year}
  {2006})}\BibitemShut {NoStop}%
\bibitem [{\citenamefont {Freire}(2008)}]{Freire2008}%
  \BibitemOpen
  \bibfield  {author} {\bibinfo {author} {\bibfnamefont {E.}~\bibnamefont
  {Freire}},\ }\bibfield  {title} {\enquote {\bibinfo {title} {Do enthalpy and
  entropy distinguish first in class from best in class?}}\ }\href {\doibase
  10.1016/j.drudis.2008.07.005} {\bibfield  {journal} {\bibinfo  {journal}
  {Drug discovery today}\ }\textbf {\bibinfo {volume} {13}},\ \bibinfo {pages}
  {869--874} (\bibinfo {year} {2008})},\ \bibinfo {note}
  {18703160[pmid]}\BibitemShut {NoStop}%
\bibitem [{\citenamefont {Ladbury}, \citenamefont {Klebe},\ and\ \citenamefont
  {Freire}(2010)}]{Ladbury2010}%
  \BibitemOpen
  \bibfield  {author} {\bibinfo {author} {\bibfnamefont {J.~E.}\ \bibnamefont
  {Ladbury}}, \bibinfo {author} {\bibfnamefont {G.}~\bibnamefont {Klebe}}, \
  and\ \bibinfo {author} {\bibfnamefont {E.}~\bibnamefont {Freire}},\
  }\bibfield  {title} {\enquote {\bibinfo {title} {Adding calorimetric data to
  decision making in lead discovery: a hot tip},}\ }\href {\doibase
  10.1038/nrd3054} {\bibfield  {journal} {\bibinfo  {journal} {Nature Reviews
  Drug Discovery}\ }\textbf {\bibinfo {volume} {9}},\ \bibinfo {pages} {23--27}
  (\bibinfo {year} {2010})}\BibitemShut {NoStop}%
\bibitem [{\citenamefont {Jen-Jacobson}, \citenamefont {Engler},\ and\
  \citenamefont {Jacobson}(2000)}]{JenJacobson2000}%
  \BibitemOpen
  \bibfield  {author} {\bibinfo {author} {\bibfnamefont {L.}~\bibnamefont
  {Jen-Jacobson}}, \bibinfo {author} {\bibfnamefont {L.~E.}\ \bibnamefont
  {Engler}}, \ and\ \bibinfo {author} {\bibfnamefont {L.~A.}\ \bibnamefont
  {Jacobson}},\ }\bibfield  {title} {\enquote {\bibinfo {title} {Structural and
  thermodynamic strategies for site-specific dna binding proteins},}\ }\href
  {\doibase https://doi.org/10.1016/S0969-2126(00)00501-3} {\bibfield
  {journal} {\bibinfo  {journal} {Structure}\ }\textbf {\bibinfo {volume}
  {8}},\ \bibinfo {pages} {1015--1023} (\bibinfo {year} {2000})}\BibitemShut
  {NoStop}%
\bibitem [{\citenamefont {Starikov}\ and\ \citenamefont
  {Nordén}(2012)}]{Starikov2012}%
  \BibitemOpen
  \bibfield  {author} {\bibinfo {author} {\bibfnamefont {E.}~\bibnamefont
  {Starikov}}\ and\ \bibinfo {author} {\bibfnamefont {B.}~\bibnamefont
  {Norden}},\ }\bibfield  {title} {\enquote {\bibinfo {title}
  {Entropy-enthalpy compensation as a fundamental concept and analysis tool for
  systematical experimental data},}\ }\href {\doibase
  10.1016/j.cplett.2012.04.028} {\bibfield  {journal} {\bibinfo  {journal}
  {Chemical Physics Letters}\ }\textbf {\bibinfo {volume} {538}},\ \bibinfo
  {pages} {118--120} (\bibinfo {year} {2012})}\BibitemShut {NoStop}%
\bibitem [{\citenamefont {Black}(2007)}]{Black2007}%
  \BibitemOpen
  \bibfield  {author} {\bibinfo {author} {\bibfnamefont {S.}~\bibnamefont
  {Black}},\ }\bibfield  {title} {\enquote {\bibinfo {title} {Simulating
  nucleation of molecular solids},}\ }\href@noop {} {\bibfield  {journal}
  {\bibinfo  {journal} {Proceedings of the Royal Society A: Mathematical,
  Physical and Engineering Sciences}\ }\textbf {\bibinfo {volume} {463}},\
  \bibinfo {pages} {2799--2811} (\bibinfo {year} {2007})}\BibitemShut {NoStop}%
\bibitem [{\citenamefont {Radhakrishnan}\ and\ \citenamefont
  {Trout}(2003)}]{Radhakrishnan2003}%
  \BibitemOpen
  \bibfield  {author} {\bibinfo {author} {\bibfnamefont {R.}~\bibnamefont
  {Radhakrishnan}}\ and\ \bibinfo {author} {\bibfnamefont {B.~L.}\ \bibnamefont
  {Trout}},\ }\bibfield  {title} {\enquote {\bibinfo {title} {Nucleation of
  hexagonal ice (ih) in liquid water},}\ }\href {\doibase 10.1021/ja0211252}
  {\bibfield  {journal} {\bibinfo  {journal} {Journal of the American Chemical
  Society}\ }\textbf {\bibinfo {volume} {125}},\ \bibinfo {pages} {7743--7747}
  (\bibinfo {year} {2003})}\BibitemShut {NoStop}%
\bibitem [{\citenamefont {Jo}\ \emph {et~al.}(2010)\citenamefont {Jo},
  \citenamefont {Rui}, \citenamefont {Lim}, \citenamefont {Klauda},\ and\
  \citenamefont {Im}}]{Jo2010}%
  \BibitemOpen
  \bibfield  {author} {\bibinfo {author} {\bibfnamefont {S.}~\bibnamefont
  {Jo}}, \bibinfo {author} {\bibfnamefont {H.}~\bibnamefont {Rui}}, \bibinfo
  {author} {\bibfnamefont {J.~B.}\ \bibnamefont {Lim}}, \bibinfo {author}
  {\bibfnamefont {J.~B.}\ \bibnamefont {Klauda}}, \ and\ \bibinfo {author}
  {\bibfnamefont {W.}~\bibnamefont {Im}},\ }\bibfield  {title} {\enquote
  {\bibinfo {title} {Cholesterol flip-flop: Insights from free energy
  simulation studies},}\ }\href {\doibase 10.1021/jp108166k} {\bibfield
  {journal} {\bibinfo  {journal} {The Journal of Physical Chemistry B}\
  }\textbf {\bibinfo {volume} {114}},\ \bibinfo {pages} {13342--13348}
  (\bibinfo {year} {2010})},\ \bibinfo {note} {pMID: 20923227},\ \Eprint
  {http://arxiv.org/abs/https://doi.org/10.1021/jp108166k}
  {https://doi.org/10.1021/jp108166k} \BibitemShut {NoStop}%
\bibitem [{\citenamefont {Marrink}\ and\ \citenamefont
  {Berendsen}(1994)}]{Marrink1994}%
  \BibitemOpen
  \bibfield  {author} {\bibinfo {author} {\bibfnamefont {S.-J.}\ \bibnamefont
  {Marrink}}\ and\ \bibinfo {author} {\bibfnamefont {H.~J.}\ \bibnamefont
  {Berendsen}},\ }\bibfield  {title} {\enquote {\bibinfo {title} {Simulation of
  water transport through a lipid membrane},}\ }\href@noop {} {\bibfield
  {journal} {\bibinfo  {journal} {The Journal of Physical Chemistry}\ }\textbf
  {\bibinfo {volume} {98}},\ \bibinfo {pages} {4155--4168} (\bibinfo {year}
  {1994})}\BibitemShut {NoStop}%
\bibitem [{\citenamefont {MacCallum}\ and\ \citenamefont
  {Tieleman}(2006)}]{Maccallum2006}%
  \BibitemOpen
  \bibfield  {author} {\bibinfo {author} {\bibfnamefont {J.~L.}\ \bibnamefont
  {MacCallum}}\ and\ \bibinfo {author} {\bibfnamefont {D.~P.}\ \bibnamefont
  {Tieleman}},\ }\bibfield  {title} {\enquote {\bibinfo {title} {Computer
  simulation of the distribution of hexane in a lipid bilayer: spatially
  resolved free energy, entropy, and enthalpy profiles},}\ }\href@noop {}
  {\bibfield  {journal} {\bibinfo  {journal} {Journal of the American Chemical
  Society}\ }\textbf {\bibinfo {volume} {128}},\ \bibinfo {pages} {125--130}
  (\bibinfo {year} {2006})}\BibitemShut {NoStop}%
\bibitem [{\citenamefont {Frenkel}(1993)}]{Frenkel1993}%
  \BibitemOpen
  \bibfield  {author} {\bibinfo {author} {\bibfnamefont {D.}~\bibnamefont
  {Frenkel}},\ }\enquote {\bibinfo {title} {Order through disorder:
  Entropy-driven phase transitions},}\ \ (\bibinfo {year} {1993})\ pp.\
  \bibinfo {pages} {137--148}\BibitemShut {NoStop}%
\bibitem [{\citenamefont {Frenkel}(2015)}]{Frenkel2015}%
  \BibitemOpen
  \bibfield  {author} {\bibinfo {author} {\bibfnamefont {D.}~\bibnamefont
  {Frenkel}},\ }\bibfield  {title} {\enquote {\bibinfo {title} {Order through
  entropy.}}\ }\href@noop {} {\bibfield  {journal} {\bibinfo  {journal} {Nature
  materials}\ }\textbf {\bibinfo {volume} {14 1}},\ \bibinfo {pages} {9--12}
  (\bibinfo {year} {2015})}\BibitemShut {NoStop}%
\bibitem [{\citenamefont {Lee}\ \emph {et~al.}(2019)\citenamefont {Lee},
  \citenamefont {Teich}, \citenamefont {Engel},\ and\ \citenamefont
  {Glotzer}}]{Lee2019}%
  \BibitemOpen
  \bibfield  {author} {\bibinfo {author} {\bibfnamefont {S.}~\bibnamefont
  {Lee}}, \bibinfo {author} {\bibfnamefont {E.~G.}\ \bibnamefont {Teich}},
  \bibinfo {author} {\bibfnamefont {M.}~\bibnamefont {Engel}}, \ and\ \bibinfo
  {author} {\bibfnamefont {S.~C.}\ \bibnamefont {Glotzer}},\ }\bibfield
  {title} {\enquote {\bibinfo {title} {Entropic colloidal crystallization
  pathways via fluid--fluid transitions and multidimensional prenucleation
  motifs},}\ }\href@noop {} {\bibfield  {journal} {\bibinfo  {journal}
  {Proceedings of the National Academy of Sciences}\ }\textbf {\bibinfo
  {volume} {116}},\ \bibinfo {pages} {14843--14851} (\bibinfo {year}
  {2019})}\BibitemShut {NoStop}%
\bibitem [{\citenamefont {Vo}\ and\ \citenamefont {Glotzer}(2022)}]{Vo2022}%
  \BibitemOpen
  \bibfield  {author} {\bibinfo {author} {\bibfnamefont {T.}~\bibnamefont
  {Vo}}\ and\ \bibinfo {author} {\bibfnamefont {S.~C.}\ \bibnamefont
  {Glotzer}},\ }\bibfield  {title} {\enquote {\bibinfo {title} {A theory of
  entropic bonding},}\ }\href {\doibase 10.1073/pnas.2116414119} {\bibfield
  {journal} {\bibinfo  {journal} {Proceedings of the National Academy of
  Sciences}\ }\textbf {\bibinfo {volume} {119}} (\bibinfo {year} {2022}),\
  10.1073/pnas.2116414119},\ \Eprint
  {http://arxiv.org/abs/https://www.pnas.org/content/119/4/e2116414119.full.pdf}
  {https://www.pnas.org/content/119/4/e2116414119.full.pdf} \BibitemShut
  {NoStop}%
\bibitem [{\citenamefont {H\"anggi}, \citenamefont {Talkner},\ and\
  \citenamefont {Borkovec}(1990)}]{Hanggi1990}%
  \BibitemOpen
  \bibfield  {author} {\bibinfo {author} {\bibfnamefont {P.}~\bibnamefont
  {H\"anggi}}, \bibinfo {author} {\bibfnamefont {P.}~\bibnamefont {Talkner}}, \
  and\ \bibinfo {author} {\bibfnamefont {M.}~\bibnamefont {Borkovec}},\
  }\bibfield  {title} {\enquote {\bibinfo {title} {Reaction-rate theory: fifty
  years after kramers},}\ }\href {\doibase 10.1103/RevModPhys.62.251}
  {\bibfield  {journal} {\bibinfo  {journal} {Rev. Mod. Phys.}\ }\textbf
  {\bibinfo {volume} {62}},\ \bibinfo {pages} {251--341} (\bibinfo {year}
  {1990})}\BibitemShut {NoStop}%
\bibitem [{\citenamefont {Guarnera}\ and\ \citenamefont
  {Vanden-Eijnden}(2016)}]{Guarnera2016}%
  \BibitemOpen
  \bibfield  {author} {\bibinfo {author} {\bibfnamefont {E.}~\bibnamefont
  {Guarnera}}\ and\ \bibinfo {author} {\bibfnamefont {E.}~\bibnamefont
  {Vanden-Eijnden}},\ }\bibfield  {title} {\enquote {\bibinfo {title}
  {Optimized markov state models for metastable systems},}\ }\href {\doibase
  10.1063/1.4954769} {\bibfield  {journal} {\bibinfo  {journal} {The Journal of
  Chemical Physics}\ }\textbf {\bibinfo {volume} {145}},\ \bibinfo {pages}
  {024102} (\bibinfo {year} {2016})},\ \Eprint
  {http://arxiv.org/abs/https://doi.org/10.1063/1.4954769}
  {https://doi.org/10.1063/1.4954769} \BibitemShut {NoStop}%
\bibitem [{\citenamefont {Gimondi}, \citenamefont {Tribello},\ and\
  \citenamefont {Salvalaglio}(2018)}]{Gimondi2018}%
  \BibitemOpen
  \bibfield  {author} {\bibinfo {author} {\bibfnamefont {I.}~\bibnamefont
  {Gimondi}}, \bibinfo {author} {\bibfnamefont {G.~A.}\ \bibnamefont
  {Tribello}}, \ and\ \bibinfo {author} {\bibfnamefont {M.}~\bibnamefont
  {Salvalaglio}},\ }\bibfield  {title} {\enquote {\bibinfo {title} {Building
  maps in collective variable space},}\ }\href {\doibase 10.1063/1.5027528}
  {\bibfield  {journal} {\bibinfo  {journal} {The Journal of Chemical Physics}\
  }\textbf {\bibinfo {volume} {149}},\ \bibinfo {pages} {104104} (\bibinfo
  {year} {2018})},\ \Eprint
  {http://arxiv.org/abs/https://doi.org/10.1063/1.5027528}
  {https://doi.org/10.1063/1.5027528} \BibitemShut {NoStop}%
\bibitem [{\citenamefont {Tsai}, \citenamefont {Smith},\ and\ \citenamefont
  {Tiwary}(2019)}]{tsai2019reaction}%
  \BibitemOpen
  \bibfield  {author} {\bibinfo {author} {\bibfnamefont {S.-T.}\ \bibnamefont
  {Tsai}}, \bibinfo {author} {\bibfnamefont {Z.}~\bibnamefont {Smith}}, \ and\
  \bibinfo {author} {\bibfnamefont {P.}~\bibnamefont {Tiwary}},\ }\bibfield
  {title} {\enquote {\bibinfo {title} {Reaction coordinates and rate constants
  for liquid droplet nucleation: Quantifying the interplay between driving
  force and memory},}\ }\href@noop {} {\bibfield  {journal} {\bibinfo
  {journal} {The Journal of chemical physics}\ }\textbf {\bibinfo {volume}
  {151}},\ \bibinfo {pages} {154106} (\bibinfo {year} {2019})}\BibitemShut
  {NoStop}%
\bibitem [{\citenamefont {Katiyar}\ and\ \citenamefont
  {Thompson}(2021)}]{Katiyar2021}%
  \BibitemOpen
  \bibfield  {author} {\bibinfo {author} {\bibfnamefont {A.}~\bibnamefont
  {Katiyar}}\ and\ \bibinfo {author} {\bibfnamefont {W.~H.}\ \bibnamefont
  {Thompson}},\ }\bibfield  {title} {\enquote {\bibinfo {title} {Temperature
  dependence of peptide conformational equilibria from simulations at a single
  temperature},}\ }\href {\doibase 10.1021/acs.jpca.1c00150} {\bibfield
  {journal} {\bibinfo  {journal} {The Journal of Physical Chemistry A}\
  }\textbf {\bibinfo {volume} {125}},\ \bibinfo {pages} {2374--2384} (\bibinfo
  {year} {2021})},\ \bibinfo {note} {pMID: 33720712},\ \Eprint
  {http://arxiv.org/abs/https://doi.org/10.1021/acs.jpca.1c00150}
  {https://doi.org/10.1021/acs.jpca.1c00150} \BibitemShut {NoStop}%
\bibitem [{\citenamefont {Bryngelson}\ \emph {et~al.}(1995)\citenamefont
  {Bryngelson}, \citenamefont {Onuchic}, \citenamefont {Socci},\ and\
  \citenamefont {Wolynes}}]{Bryngelson1995}%
  \BibitemOpen
  \bibfield  {author} {\bibinfo {author} {\bibfnamefont {J.~D.}\ \bibnamefont
  {Bryngelson}}, \bibinfo {author} {\bibfnamefont {J.~N.}\ \bibnamefont
  {Onuchic}}, \bibinfo {author} {\bibfnamefont {N.~D.}\ \bibnamefont {Socci}},
  \ and\ \bibinfo {author} {\bibfnamefont {P.~G.}\ \bibnamefont {Wolynes}},\
  }\bibfield  {title} {\enquote {\bibinfo {title} {Funnels, pathways, and the
  energy landscape of protein folding: A synthesis},}\ }\href {\doibase
  10.1002/prot.340210302} {\bibfield  {journal} {\bibinfo  {journal} {Proteins:
  Structure, Function, and Bioinformatics}\ }\textbf {\bibinfo {volume} {21}},\
  \bibinfo {pages} {167--195} (\bibinfo {year} {1995})},\ \Eprint
  {http://arxiv.org/abs/https://onlinelibrary.wiley.com/doi/pdf/10.1002/prot.340210302}
  {https://onlinelibrary.wiley.com/doi/pdf/10.1002/prot.340210302} \BibitemShut
  {NoStop}%
\bibitem [{\citenamefont {Onuchic}, \citenamefont {Luthey-Schulten},\ and\
  \citenamefont {Wolynes}(1997)}]{Onuchic1997}%
  \BibitemOpen
  \bibfield  {author} {\bibinfo {author} {\bibfnamefont {J.~N.}\ \bibnamefont
  {Onuchic}}, \bibinfo {author} {\bibfnamefont {Z.}~\bibnamefont
  {Luthey-Schulten}}, \ and\ \bibinfo {author} {\bibfnamefont {P.~G.}\
  \bibnamefont {Wolynes}},\ }\bibfield  {title} {\enquote {\bibinfo {title}
  {{THEORY OF PROTEIN FOLDING: The Energy Landscape Perspective}},}\ }\href
  {\doibase 10.1146/annurev.physchem.48.1.545} {\bibfield  {journal} {\bibinfo
  {journal} {Annual Review of Physical Chemistry}\ }\textbf {\bibinfo {volume}
  {48}},\ \bibinfo {pages} {545--600} (\bibinfo {year} {1997})}\BibitemShut
  {NoStop}%
\bibitem [{\citenamefont {Zhuravlev}\ and\ \citenamefont
  {Papoian}(2010)}]{Zhuravlev2010}%
  \BibitemOpen
  \bibfield  {author} {\bibinfo {author} {\bibfnamefont {P.}~\bibnamefont
  {Zhuravlev}}\ and\ \bibinfo {author} {\bibfnamefont {G.}~\bibnamefont
  {Papoian}},\ }\bibfield  {title} {\enquote {\bibinfo {title} {Protein
  functional landscapes, dynamics, allostery: a tortuous path towards a
  universal theoretical framework},}\ }\href {\doibase
  10.1017/S0033583510000119} {\bibfield  {journal} {\bibinfo  {journal}
  {Quarterly Reviews of Biophysics}\ }\textbf {\bibinfo {volume} {43}},\
  \bibinfo {pages} {295?332} (\bibinfo {year} {2010})}\BibitemShut {NoStop}%
\bibitem [{\citenamefont {Leoni}\ and\ \citenamefont
  {Russo}(2021)}]{Leoni2021}%
  \BibitemOpen
  \bibfield  {author} {\bibinfo {author} {\bibfnamefont {F.}~\bibnamefont
  {Leoni}}\ and\ \bibinfo {author} {\bibfnamefont {J.}~\bibnamefont {Russo}},\
  }\bibfield  {title} {\enquote {\bibinfo {title} {Nonclassical nucleation
  pathways in stacking-disordered crystals},}\ }\href {\doibase
  10.1103/PhysRevX.11.031006} {\bibfield  {journal} {\bibinfo  {journal} {Phys.
  Rev. X}\ }\textbf {\bibinfo {volume} {11}},\ \bibinfo {pages} {031006}
  (\bibinfo {year} {2021})}\BibitemShut {NoStop}%
\bibitem [{\citenamefont {Jiang}\ \emph {et~al.}(2018)\citenamefont {Jiang},
  \citenamefont {Haji-Akbari}, \citenamefont {Debenedetti},\ and\ \citenamefont
  {Panagiotopoulos}}]{Jiang2018}%
  \BibitemOpen
  \bibfield  {author} {\bibinfo {author} {\bibfnamefont {H.}~\bibnamefont
  {Jiang}}, \bibinfo {author} {\bibfnamefont {A.}~\bibnamefont {Haji-Akbari}},
  \bibinfo {author} {\bibfnamefont {P.~G.}\ \bibnamefont {Debenedetti}}, \ and\
  \bibinfo {author} {\bibfnamefont {A.~Z.}\ \bibnamefont {Panagiotopoulos}},\
  }\bibfield  {title} {\enquote {\bibinfo {title} {Forward flux sampling
  calculation of homogeneous nucleation rates from aqueous nacl solutions},}\
  }\href {\doibase 10.1063/1.5016554} {\bibfield  {journal} {\bibinfo
  {journal} {The Journal of Chemical Physics}\ }\textbf {\bibinfo {volume}
  {148}},\ \bibinfo {pages} {044505} (\bibinfo {year} {2018})},\ \Eprint
  {http://arxiv.org/abs/https://doi.org/10.1063/1.5016554}
  {https://doi.org/10.1063/1.5016554} \BibitemShut {NoStop}%
\bibitem [{\citenamefont {Jiang}, \citenamefont {Debenedetti},\ and\
  \citenamefont {Panagiotopoulos}(2019)}]{Jiang2019}%
  \BibitemOpen
  \bibfield  {author} {\bibinfo {author} {\bibfnamefont {H.}~\bibnamefont
  {Jiang}}, \bibinfo {author} {\bibfnamefont {P.~G.}\ \bibnamefont
  {Debenedetti}}, \ and\ \bibinfo {author} {\bibfnamefont {A.~Z.}\ \bibnamefont
  {Panagiotopoulos}},\ }\bibfield  {title} {\enquote {\bibinfo {title}
  {Nucleation in aqueous nacl solutions shifts from 1-step to 2-step mechanism
  on crossing the spinodal},}\ }\href {\doibase 10.1063/1.5084248} {\bibfield
  {journal} {\bibinfo  {journal} {The Journal of Chemical Physics}\ }\textbf
  {\bibinfo {volume} {150}},\ \bibinfo {pages} {124502} (\bibinfo {year}
  {2019})},\ \Eprint {http://arxiv.org/abs/https://doi.org/10.1063/1.5084248}
  {https://doi.org/10.1063/1.5084248} \BibitemShut {NoStop}%
\bibitem [{\citenamefont {Schwantes}\ and\ \citenamefont
  {Pande}(2013)}]{Schwantes2013}%
  \BibitemOpen
  \bibfield  {author} {\bibinfo {author} {\bibfnamefont {C.~R.}\ \bibnamefont
  {Schwantes}}\ and\ \bibinfo {author} {\bibfnamefont {V.~S.}\ \bibnamefont
  {Pande}},\ }\bibfield  {title} {\enquote {\bibinfo {title} {Improvements in
  markov state model construction reveal many non-native interactions in the
  folding of ntl9},}\ }\href {\doibase 10.1021/ct300878a} {\bibfield  {journal}
  {\bibinfo  {journal} {Journal of Chemical Theory and Computation}\ }\textbf
  {\bibinfo {volume} {9}},\ \bibinfo {pages} {2000--2009} (\bibinfo {year}
  {2013})},\ \bibinfo {note} {pMID: 23750122},\ \Eprint
  {http://arxiv.org/abs/https://doi.org/10.1021/ct300878a}
  {https://doi.org/10.1021/ct300878a} \BibitemShut {NoStop}%
\bibitem [{\citenamefont {Beauchamp}\ \emph {et~al.}(2012)\citenamefont
  {Beauchamp}, \citenamefont {McGibbon}, \citenamefont {Lin},\ and\
  \citenamefont {Pande}}]{Beauchamp2012}%
  \BibitemOpen
  \bibfield  {author} {\bibinfo {author} {\bibfnamefont {K.~A.}\ \bibnamefont
  {Beauchamp}}, \bibinfo {author} {\bibfnamefont {R.}~\bibnamefont {McGibbon}},
  \bibinfo {author} {\bibfnamefont {Y.-S.}\ \bibnamefont {Lin}}, \ and\
  \bibinfo {author} {\bibfnamefont {V.~S.}\ \bibnamefont {Pande}},\ }\bibfield
  {title} {\enquote {\bibinfo {title} {Simple few-state models reveal hidden
  complexity in protein folding},}\ }\href {\doibase 10.1073/pnas.1201810109}
  {\bibfield  {journal} {\bibinfo  {journal} {Proceedings of the National
  Academy of Sciences}\ }\textbf {\bibinfo {volume} {109}},\ \bibinfo {pages}
  {17807--17813} (\bibinfo {year} {2012})},\ \Eprint
  {http://arxiv.org/abs/https://www.pnas.org/content/109/44/17807.full.pdf}
  {https://www.pnas.org/content/109/44/17807.full.pdf} \BibitemShut {NoStop}%
\bibitem [{\citenamefont {Tsai}, \citenamefont {Smith},\ and\ \citenamefont
  {Tiwary}(2021)}]{tsai2021sgoop}%
  \BibitemOpen
  \bibfield  {author} {\bibinfo {author} {\bibfnamefont {S.-T.}\ \bibnamefont
  {Tsai}}, \bibinfo {author} {\bibfnamefont {Z.}~\bibnamefont {Smith}}, \ and\
  \bibinfo {author} {\bibfnamefont {P.}~\bibnamefont {Tiwary}},\ }\bibfield
  {title} {\enquote {\bibinfo {title} {Sgoop-d: Estimating kinetic distances
  and reaction coordinate dimensionality for rare event systems from
  biased/unbiased simulations},}\ }\href@noop {} {\bibfield  {journal}
  {\bibinfo  {journal} {Journal of Chemical Theory and Computation}\ }\textbf
  {\bibinfo {volume} {17}},\ \bibinfo {pages} {6757--6765} (\bibinfo {year}
  {2021})}\BibitemShut {NoStop}%
\bibitem [{\citenamefont {Finney}\ and\ \citenamefont
  {Salvalaglio}(2021)}]{Finney2021}%
  \BibitemOpen
  \bibfield  {author} {\bibinfo {author} {\bibfnamefont {A.}~\bibnamefont
  {Finney}}\ and\ \bibinfo {author} {\bibfnamefont {M.}~\bibnamefont
  {Salvalaglio}},\ }\bibfield  {title} {\enquote {\bibinfo {title} {Multiple
  pathways in nacl homogeneous crystal nucleation},}\ }\href {\doibase
  10.1039/D1FD00089F} {\bibfield  {journal} {\bibinfo  {journal} {Faraday
  Discuss.}\ ,\ \bibinfo {pages} {--}} (\bibinfo {year} {2021})}\BibitemShut
  {NoStop}%
\bibitem [{\citenamefont {Tiwary}\ and\ \citenamefont {van~de
  Walle}(2016)}]{Tiwary2016}%
  \BibitemOpen
  \bibfield  {author} {\bibinfo {author} {\bibfnamefont {P.}~\bibnamefont
  {Tiwary}}\ and\ \bibinfo {author} {\bibfnamefont {A.}~\bibnamefont {van~de
  Walle}},\ }\enquote {\bibinfo {title} {A review of enhanced sampling
  approaches for accelerated molecular dynamics},}\ in\ \href {\doibase
  10.1007/978-3-319-33480-6_6} {\emph {\bibinfo {booktitle} {Multiscale
  Materials Modeling for Nanomechanics}}},\ \bibinfo {editor} {edited by\
  \bibinfo {editor} {\bibfnamefont {C.~R.}\ \bibnamefont {Weinberger}}\ and\
  \bibinfo {editor} {\bibfnamefont {G.~J.}\ \bibnamefont {Tucker}}}\ (\bibinfo
  {publisher} {Springer International Publishing},\ \bibinfo {address} {Cham},\
  \bibinfo {year} {2016})\ pp.\ \bibinfo {pages} {195--221}\BibitemShut
  {NoStop}%
\bibitem [{\citenamefont {Tiwary}\ and\ \citenamefont {van~de
  Walle}(2013)}]{tiwary2013accelerated}%
  \BibitemOpen
  \bibfield  {author} {\bibinfo {author} {\bibfnamefont {P.}~\bibnamefont
  {Tiwary}}\ and\ \bibinfo {author} {\bibfnamefont {A.}~\bibnamefont {van~de
  Walle}},\ }\bibfield  {title} {\enquote {\bibinfo {title} {Accelerated
  molecular dynamics through stochastic iterations and collective variable
  based basin identification},}\ }\href@noop {} {\bibfield  {journal} {\bibinfo
   {journal} {Physical Review B}\ }\textbf {\bibinfo {volume} {87}},\ \bibinfo
  {pages} {094304} (\bibinfo {year} {2013})}\BibitemShut {NoStop}%
\bibitem [{\citenamefont {Bussi}\ and\ \citenamefont {Laio}(2020)}]{Bussi2020}%
  \BibitemOpen
  \bibfield  {author} {\bibinfo {author} {\bibfnamefont {G.}~\bibnamefont
  {Bussi}}\ and\ \bibinfo {author} {\bibfnamefont {A.}~\bibnamefont {Laio}},\
  }\bibfield  {title} {\enquote {\bibinfo {title} {Using metadynamics to
  explore complex free-energy landscapes},}\ }\href {\doibase
  10.1038/s42254-020-0153-0} {\bibfield  {journal} {\bibinfo  {journal} {Nature
  Reviews Physics}\ }\textbf {\bibinfo {volume} {2}},\ \bibinfo {pages}
  {200--212} (\bibinfo {year} {2020})}\BibitemShut {NoStop}%
\bibitem [{\citenamefont {Hyv{\"a}rinen}, \citenamefont {Karhunen},\ and\
  \citenamefont {Oja}(2001)}]{Hyvarinen2001}%
  \BibitemOpen
  \bibfield  {author} {\bibinfo {author} {\bibfnamefont {A.}~\bibnamefont
  {Hyv{\"a}rinen}}, \bibinfo {author} {\bibfnamefont {J.}~\bibnamefont
  {Karhunen}}, \ and\ \bibinfo {author} {\bibfnamefont {E.}~\bibnamefont
  {Oja}},\ }\href {https://books.google.com/books?id=9TQNEAAAQBAJ} {\emph
  {\bibinfo {title} {Independent Component Analysis}}},\ Adaptive and Cognitive
  Dynamic Systems: Signal Processing, Learning, Communications and Control\
  (\bibinfo  {publisher} {Wiley},\ \bibinfo {year} {2001})\BibitemShut
  {NoStop}%
\bibitem [{\citenamefont {Bowman}, \citenamefont {Pande},\ and\ \citenamefont
  {No{\'{e}}}(2013)}]{Bowman2013}%
  \BibitemOpen
  \bibfield  {author} {\bibinfo {author} {\bibfnamefont {G.~R.}\ \bibnamefont
  {Bowman}}, \bibinfo {author} {\bibfnamefont {V.~S.}\ \bibnamefont {Pande}}, \
  and\ \bibinfo {author} {\bibfnamefont {F.}~\bibnamefont {No{\'{e}}}},\
  }\href@noop {} {\emph {\bibinfo {title} {{An Introduction to Markov State
  Models and Their Application to Long Timescale Molecular Simulation (Advances
  in Experimental Medicine and Biology}}}}\ (\bibinfo {year}
  {2013})\BibitemShut {NoStop}%
\bibitem [{\citenamefont {Valsson}, \citenamefont {Tiwary},\ and\ \citenamefont
  {Parrinello}(2016)}]{Valsson2016}%
  \BibitemOpen
  \bibfield  {author} {\bibinfo {author} {\bibfnamefont {O.}~\bibnamefont
  {Valsson}}, \bibinfo {author} {\bibfnamefont {P.}~\bibnamefont {Tiwary}}, \
  and\ \bibinfo {author} {\bibfnamefont {M.}~\bibnamefont {Parrinello}},\
  }\bibfield  {title} {\enquote {\bibinfo {title} {Enhancing important
  fluctuations: Rare events and metadynamics from a conceptual viewpoint},}\
  }\href {\doibase 10.1146/annurev-physchem-040215-112229} {\bibfield
  {journal} {\bibinfo  {journal} {Annual Review of Physical Chemistry}\
  }\textbf {\bibinfo {volume} {67}},\ \bibinfo {pages} {159--184} (\bibinfo
  {year} {2016})},\ \bibinfo {note} {pMID: 26980304},\ \Eprint
  {http://arxiv.org/abs/https://doi.org/10.1146/annurev-physchem-040215-112229}
  {https://doi.org/10.1146/annurev-physchem-040215-112229} \BibitemShut
  {NoStop}%
\bibitem [{\citenamefont {Wang}\ and\ \citenamefont
  {Tiwary}(2021)}]{Wang2021a}%
  \BibitemOpen
  \bibfield  {author} {\bibinfo {author} {\bibfnamefont {D.}~\bibnamefont
  {Wang}}\ and\ \bibinfo {author} {\bibfnamefont {P.}~\bibnamefont {Tiwary}},\
  }\bibfield  {title} {\enquote {\bibinfo {title} {{State predictive
  information bottleneck}},}\ }\href {\doibase 10.1063/5.0038198} {\bibfield
  {journal} {\bibinfo  {journal} {Journal of Chemical Physics}\ }\textbf
  {\bibinfo {volume} {154}} (\bibinfo {year} {2021}),\ 10.1063/5.0038198},\
  \Eprint {http://arxiv.org/abs/2011.10127} {arXiv:2011.10127} \BibitemShut
  {NoStop}%
\bibitem [{\citenamefont {Alemi}\ \emph {et~al.}(2017)\citenamefont {Alemi},
  \citenamefont {Fischer}, \citenamefont {Dillon},\ and\ \citenamefont
  {Murphy}}]{Alemi2017}%
  \BibitemOpen
  \bibfield  {author} {\bibinfo {author} {\bibfnamefont {A.~A.}\ \bibnamefont
  {Alemi}}, \bibinfo {author} {\bibfnamefont {I.}~\bibnamefont {Fischer}},
  \bibinfo {author} {\bibfnamefont {J.~V.}\ \bibnamefont {Dillon}}, \ and\
  \bibinfo {author} {\bibfnamefont {K.}~\bibnamefont {Murphy}},\ }\bibfield
  {title} {\enquote {\bibinfo {title} {{Deep variational information
  bottleneck}},}\ }\href@noop {} {\bibfield  {journal} {\bibinfo  {journal}
  {5th International Conference on Learning Representations, ICLR 2017 -
  Conference Track Proceedings}\ ,\ \bibinfo {pages} {1--19}} (\bibinfo {year}
  {2017})},\ \Eprint {http://arxiv.org/abs/1612.00410} {arXiv:1612.00410}
  \BibitemShut {NoStop}%
\bibitem [{\citenamefont {Laio}\ and\ \citenamefont
  {Parrinello}(2002)}]{Laio2002}%
  \BibitemOpen
  \bibfield  {author} {\bibinfo {author} {\bibfnamefont {A.}~\bibnamefont
  {Laio}}\ and\ \bibinfo {author} {\bibfnamefont {M.}~\bibnamefont
  {Parrinello}},\ }\bibfield  {title} {\enquote {\bibinfo {title} {Escaping
  free-energy minima},}\ }\href {\doibase 10.1073/pnas.202427399} {\bibfield
  {journal} {\bibinfo  {journal} {Proceedings of the National Academy of
  Sciences}\ }\textbf {\bibinfo {volume} {99}},\ \bibinfo {pages}
  {12562--12566} (\bibinfo {year} {2002})},\ \Eprint
  {http://arxiv.org/abs/https://www.pnas.org/content/99/20/12562.full.pdf}
  {https://www.pnas.org/content/99/20/12562.full.pdf} \BibitemShut {NoStop}%
\bibitem [{\citenamefont {Barducci}, \citenamefont {Bussi},\ and\ \citenamefont
  {Parrinello}(2008)}]{Barducci2007}%
  \BibitemOpen
  \bibfield  {author} {\bibinfo {author} {\bibfnamefont {A.}~\bibnamefont
  {Barducci}}, \bibinfo {author} {\bibfnamefont {G.}~\bibnamefont {Bussi}}, \
  and\ \bibinfo {author} {\bibfnamefont {M.}~\bibnamefont {Parrinello}},\
  }\bibfield  {title} {\enquote {\bibinfo {title} {Well-tempered metadynamics:
  A smoothly converging and tunable free-energy method},}\ }\href {\doibase
  10.1103/PhysRevLett.100.020603} {\bibfield  {journal} {\bibinfo  {journal}
  {Phys. Rev. Lett.}\ }\textbf {\bibinfo {volume} {100}},\ \bibinfo {pages}
  {020603} (\bibinfo {year} {2008})}\BibitemShut {NoStop}%
\bibitem [{\citenamefont {Torrie}\ and\ \citenamefont
  {Valleau}(1977)}]{Torrie1977}%
  \BibitemOpen
  \bibfield  {author} {\bibinfo {author} {\bibfnamefont {G.~M.}\ \bibnamefont
  {Torrie}}\ and\ \bibinfo {author} {\bibfnamefont {J.~P.}\ \bibnamefont
  {Valleau}},\ }\bibfield  {title} {\enquote {\bibinfo {title} {{Nonphysical
  sampling distributions in Monte Carlo free-energy estimation: Umbrella
  sampling}},}\ }\href {\doibase 10.1016/0021-9991(77)90121-8} {\bibfield
  {journal} {\bibinfo  {journal} {Journal of Computational Physics}\ }\textbf
  {\bibinfo {volume} {23}},\ \bibinfo {pages} {187--199} (\bibinfo {year}
  {1977})},\ \Eprint {http://arxiv.org/abs/NIHMS150003} {arXiv:NIHMS150003}
  \BibitemShut {NoStop}%
\bibitem [{\citenamefont {Ribeiro}\ \emph {et~al.}(2018)\citenamefont
  {Ribeiro}, \citenamefont {Bravo}, \citenamefont {Wang},\ and\ \citenamefont
  {Tiwary}}]{Ribeiro2018}%
  \BibitemOpen
  \bibfield  {author} {\bibinfo {author} {\bibfnamefont {J.~M.~L.}\
  \bibnamefont {Ribeiro}}, \bibinfo {author} {\bibfnamefont {P.}~\bibnamefont
  {Bravo}}, \bibinfo {author} {\bibfnamefont {Y.}~\bibnamefont {Wang}}, \ and\
  \bibinfo {author} {\bibfnamefont {P.}~\bibnamefont {Tiwary}},\ }\bibfield
  {title} {\enquote {\bibinfo {title} {Reweighted autoencoded variational bayes
  for enhanced sampling (rave)},}\ }\href {\doibase 10.1063/1.5025487}
  {\bibfield  {journal} {\bibinfo  {journal} {The Journal of Chemical Physics}\
  }\textbf {\bibinfo {volume} {149}},\ \bibinfo {pages} {072301} (\bibinfo
  {year} {2018})},\ \Eprint
  {http://arxiv.org/abs/https://doi.org/10.1063/1.5025487}
  {https://doi.org/10.1063/1.5025487} \BibitemShut {NoStop}%
\bibitem [{\citenamefont {Pant}\ \emph {et~al.}(2020)\citenamefont {Pant},
  \citenamefont {Smith}, \citenamefont {Wang}, \citenamefont {Tajkhorshid},\
  and\ \citenamefont {Tiwary}}]{Pant2020}%
  \BibitemOpen
  \bibfield  {author} {\bibinfo {author} {\bibfnamefont {S.}~\bibnamefont
  {Pant}}, \bibinfo {author} {\bibfnamefont {Z.}~\bibnamefont {Smith}},
  \bibinfo {author} {\bibfnamefont {Y.}~\bibnamefont {Wang}}, \bibinfo {author}
  {\bibfnamefont {E.}~\bibnamefont {Tajkhorshid}}, \ and\ \bibinfo {author}
  {\bibfnamefont {P.}~\bibnamefont {Tiwary}},\ }\bibfield  {title} {\enquote
  {\bibinfo {title} {Confronting pitfalls of ai-augmented molecular dynamics
  using statistical physics},}\ }\href {\doibase 10.1063/5.0030931} {\bibfield
  {journal} {\bibinfo  {journal} {The Journal of Chemical Physics}\ }\textbf
  {\bibinfo {volume} {153}},\ \bibinfo {pages} {234118} (\bibinfo {year}
  {2020})},\ \Eprint {http://arxiv.org/abs/https://doi.org/10.1063/5.0030931}
  {https://doi.org/10.1063/5.0030931} \BibitemShut {NoStop}%
\bibitem [{\citenamefont {Mehdi}\ \emph {et~al.}(2021)\citenamefont {Mehdi},
  \citenamefont {Wang}, \citenamefont {Pant},\ and\ \citenamefont
  {Tiwary}}]{Mehdi2021}%
  \BibitemOpen
  \bibfield  {author} {\bibinfo {author} {\bibfnamefont {S.}~\bibnamefont
  {Mehdi}}, \bibinfo {author} {\bibfnamefont {D.}~\bibnamefont {Wang}},
  \bibinfo {author} {\bibfnamefont {S.}~\bibnamefont {Pant}}, \ and\ \bibinfo
  {author} {\bibfnamefont {P.}~\bibnamefont {Tiwary}},\ }\href@noop {}
  {\enquote {\bibinfo {title} {Accelerating all-atom simulations and gaining
  mechanistic understanding of biophysical systems through state predictive
  information bottleneck},}\ } (\bibinfo {year} {2021}),\ \Eprint
  {http://arxiv.org/abs/2112.11201} {arXiv:2112.11201 [physics.bio-ph]}
  \BibitemShut {NoStop}%
\bibitem [{\citenamefont {Wang}\ and\ \citenamefont {Tiwary}(2020)}]{Wang2020}%
  \BibitemOpen
  \bibfield  {author} {\bibinfo {author} {\bibfnamefont {Y.}~\bibnamefont
  {Wang}}\ and\ \bibinfo {author} {\bibfnamefont {P.}~\bibnamefont {Tiwary}},\
  }\bibfield  {title} {\enquote {\bibinfo {title} {Understanding the role of
  predictive time delay and biased propagator in rave},}\ }\href@noop {}
  {\bibfield  {journal} {\bibinfo  {journal} {The Journal of chemical physics}\
  }\textbf {\bibinfo {volume} {152}},\ \bibinfo {pages} {144102} (\bibinfo
  {year} {2020})}\BibitemShut {NoStop}%
\bibitem [{\citenamefont {Kingma}\ and\ \citenamefont
  {Welling}(2014)}]{Kingma2014}%
  \BibitemOpen
  \bibfield  {author} {\bibinfo {author} {\bibfnamefont {D.~P.}\ \bibnamefont
  {Kingma}}\ and\ \bibinfo {author} {\bibfnamefont {M.}~\bibnamefont
  {Welling}},\ }\bibfield  {title} {\enquote {\bibinfo {title} {{Auto-encoding
  variational bayes}},}\ }\href@noop {} {\bibfield  {journal} {\bibinfo
  {journal} {2nd International Conference on Learning Representations, ICLR
  2014 - Conference Track Proceedings}\ ,\ \bibinfo {pages} {1--14}} (\bibinfo
  {year} {2014})},\ \Eprint {http://arxiv.org/abs/1312.6114} {arXiv:1312.6114}
  \BibitemShut {NoStop}%
\bibitem [{\citenamefont {Goodfellow}, \citenamefont {Bengio},\ and\
  \citenamefont {Courville}(2016)}]{Goodfellow2016}%
  \BibitemOpen
  \bibfield  {author} {\bibinfo {author} {\bibfnamefont {I.}~\bibnamefont
  {Goodfellow}}, \bibinfo {author} {\bibfnamefont {Y.}~\bibnamefont {Bengio}},
  \ and\ \bibinfo {author} {\bibfnamefont {A.}~\bibnamefont {Courville}},\
  }\href@noop {} {\emph {\bibinfo {title} {Deep Learning}}}\ (\bibinfo
  {publisher} {MIT Press},\ \bibinfo {year} {2016})\BibitemShut {NoStop}%
\bibitem [{\citenamefont {Tomczak}\ and\ \citenamefont
  {Welling}(2018)}]{Tomczak2018}%
  \BibitemOpen
  \bibfield  {author} {\bibinfo {author} {\bibfnamefont {J.~M.}\ \bibnamefont
  {Tomczak}}\ and\ \bibinfo {author} {\bibfnamefont {M.}~\bibnamefont
  {Welling}},\ }\bibfield  {title} {\enquote {\bibinfo {title} {{VAE with a
  vampprior}},}\ }\href@noop {} {\bibfield  {journal} {\bibinfo  {journal}
  {International Conference on Artificial Intelligence and Statistics, AISTATS
  2018}\ ,\ \bibinfo {pages} {1214--1223}} (\bibinfo {year} {2018})},\ \Eprint
  {http://arxiv.org/abs/1705.07120} {arXiv:1705.07120} \BibitemShut {NoStop}%
\bibitem [{\citenamefont {Deuflhard}\ and\ \citenamefont
  {Weber}(2005)}]{Deuflhard2005}%
  \BibitemOpen
  \bibfield  {author} {\bibinfo {author} {\bibfnamefont {P.}~\bibnamefont
  {Deuflhard}}\ and\ \bibinfo {author} {\bibfnamefont {M.}~\bibnamefont
  {Weber}},\ }\bibfield  {title} {\enquote {\bibinfo {title} {{Robust Perron
  cluster analysis in conformation dynamics}},}\ }\href {\doibase
  10.1016/j.laa.2004.10.026} {\bibfield  {journal} {\bibinfo  {journal} {Linear
  Algebra and Its Applications}\ }\textbf {\bibinfo {volume} {398}},\ \bibinfo
  {pages} {161--184} (\bibinfo {year} {2005})}\BibitemShut {NoStop}%
\bibitem [{\citenamefont {Kollias}\ \emph {et~al.}(2020)\citenamefont
  {Kollias}, \citenamefont {Cantu}, \citenamefont {Glezakou}, \citenamefont
  {Rousseau},\ and\ \citenamefont {Salvalaglio}}]{Kollias2020}%
  \BibitemOpen
  \bibfield  {author} {\bibinfo {author} {\bibfnamefont {L.}~\bibnamefont
  {Kollias}}, \bibinfo {author} {\bibfnamefont {D.~C.}\ \bibnamefont {Cantu}},
  \bibinfo {author} {\bibfnamefont {V.-A.}\ \bibnamefont {Glezakou}}, \bibinfo
  {author} {\bibfnamefont {R.}~\bibnamefont {Rousseau}}, \ and\ \bibinfo
  {author} {\bibfnamefont {M.}~\bibnamefont {Salvalaglio}},\ }\bibfield
  {title} {\enquote {\bibinfo {title} {On the role of enthalpic and entropic
  contributions to the conformational free energy landscape of mil-101(cr)
  secondary building units},}\ }\href {\doibase
  https://doi.org/10.1002/adts.202000092} {\bibfield  {journal} {\bibinfo
  {journal} {Advanced Theory and Simulations}\ }\textbf {\bibinfo {volume}
  {3}},\ \bibinfo {pages} {2000092} (\bibinfo {year} {2020})},\ \Eprint
  {http://arxiv.org/abs/https://onlinelibrary.wiley.com/doi/pdf/10.1002/adts.202000092}
  {https://onlinelibrary.wiley.com/doi/pdf/10.1002/adts.202000092} \BibitemShut
  {NoStop}%
\bibitem [{\citenamefont {Silverman}(1986)}]{Silverman1986}%
  \BibitemOpen
  \bibfield  {author} {\bibinfo {author} {\bibfnamefont {B.}~\bibnamefont
  {Silverman}},\ }\href {https://books.google.com/books?id=e-xsrjsL7WkC} {\emph
  {\bibinfo {title} {Density Estimation for Statistics and Data Analysis}}},\
  Chapman \& Hall/CRC Monographs on Statistics \& Applied Probability\
  (\bibinfo  {publisher} {Taylor \& Francis},\ \bibinfo {year}
  {1986})\BibitemShut {NoStop}%
\bibitem [{\citenamefont {Hartmann}, \citenamefont {Latorre},\ and\
  \citenamefont {Ciccotti}(2011)}]{Hartmann2011}%
  \BibitemOpen
  \bibfield  {author} {\bibinfo {author} {\bibfnamefont {C.}~\bibnamefont
  {Hartmann}}, \bibinfo {author} {\bibfnamefont {J.~C.}\ \bibnamefont
  {Latorre}}, \ and\ \bibinfo {author} {\bibfnamefont {G.}~\bibnamefont
  {Ciccotti}},\ }\bibfield  {title} {\enquote {\bibinfo {title} {{On two
  possible definitions of the free energy for collective variables}},}\ }\href
  {\doibase 10.1140/epjst/e2011-01519-7} {\bibfield  {journal} {\bibinfo
  {journal} {European Physical Journal: Special Topics}\ }\textbf {\bibinfo
  {volume} {200}},\ \bibinfo {pages} {73--89} (\bibinfo {year}
  {2011})}\BibitemShut {NoStop}%
\bibitem [{\citenamefont {Lelievre}, \citenamefont {Rousset},\ and\
  \citenamefont {Stoltz}(2010)}]{Lelievre2010}%
  \BibitemOpen
  \bibfield  {author} {\bibinfo {author} {\bibfnamefont {T.}~\bibnamefont
  {Lelievre}}, \bibinfo {author} {\bibfnamefont {M.}~\bibnamefont {Rousset}}, \
  and\ \bibinfo {author} {\bibfnamefont {G.}~\bibnamefont {Stoltz}},\ }\href
  {https://books.google.com/books?id=DjrICgAAQBAJ} {\emph {\bibinfo {title}
  {Free Energy Computations: A Mathematical Perspective}}}\ (\bibinfo
  {publisher} {World Scientific Publishing Company},\ \bibinfo {year}
  {2010})\BibitemShut {NoStop}%
\bibitem [{\citenamefont {Faradjian}\ and\ \citenamefont
  {Elber}(2004)}]{Faradjian2004}%
  \BibitemOpen
  \bibfield  {author} {\bibinfo {author} {\bibfnamefont {A.~K.}\ \bibnamefont
  {Faradjian}}\ and\ \bibinfo {author} {\bibfnamefont {R.}~\bibnamefont
  {Elber}},\ }\bibfield  {title} {\enquote {\bibinfo {title} {{Computing time
  scales from reaction coordinates by milestoning}},}\ }\href {\doibase
  10.1063/1.1738640} {\bibfield  {journal} {\bibinfo  {journal} {Journal of
  Chemical Physics}\ }\textbf {\bibinfo {volume} {120}},\ \bibinfo {pages}
  {10880--10889} (\bibinfo {year} {2004})},\ \Eprint
  {http://arxiv.org/abs/arXiv:1011.1669v3} {arXiv:arXiv:1011.1669v3}
  \BibitemShut {NoStop}%
\bibitem [{\citenamefont {Banisch}\ \emph {et~al.}(2020)\citenamefont
  {Banisch}, \citenamefont {Trstanova}, \citenamefont {Bittracher},
  \citenamefont {Klus},\ and\ \citenamefont {Koltai}}]{Banisch2020}%
  \BibitemOpen
  \bibfield  {author} {\bibinfo {author} {\bibfnamefont {R.}~\bibnamefont
  {Banisch}}, \bibinfo {author} {\bibfnamefont {Z.}~\bibnamefont {Trstanova}},
  \bibinfo {author} {\bibfnamefont {A.}~\bibnamefont {Bittracher}}, \bibinfo
  {author} {\bibfnamefont {S.}~\bibnamefont {Klus}}, \ and\ \bibinfo {author}
  {\bibfnamefont {P.}~\bibnamefont {Koltai}},\ }\bibfield  {title} {\enquote
  {\bibinfo {title} {{Diffusion maps tailored to arbitrary non-degenerate
  It{\^{o}} processes}},}\ }\href {\doibase 10.1016/j.acha.2018.05.001}
  {\bibfield  {journal} {\bibinfo  {journal} {Applied and Computational
  Harmonic Analysis}\ }\textbf {\bibinfo {volume} {48}},\ \bibinfo {pages}
  {242--265} (\bibinfo {year} {2020})},\ \Eprint
  {http://arxiv.org/abs/1710.03484} {arXiv:1710.03484} \BibitemShut {NoStop}%
\bibitem [{\citenamefont {Shinoda}(2016)}]{Shinoda2016}%
  \BibitemOpen
  \bibfield  {author} {\bibinfo {author} {\bibfnamefont {W.}~\bibnamefont
  {Shinoda}},\ }\bibfield  {title} {\enquote {\bibinfo {title} {{Permeability
  across lipid membranes}},}\ }\href {\doibase 10.1016/j.bbamem.2016.03.032}
  {\bibfield  {journal} {\bibinfo  {journal} {Biochimica et Biophysica Acta -
  Biomembranes}\ }\textbf {\bibinfo {volume} {1858}},\ \bibinfo {pages}
  {2254--2265} (\bibinfo {year} {2016})}\BibitemShut {NoStop}%
\bibitem [{\citenamefont {Lee}\ \emph {et~al.}(2016)\citenamefont {Lee},
  \citenamefont {Comer}, \citenamefont {Herndon}, \citenamefont {Leung},
  \citenamefont {Pavlova}, \citenamefont {Swift}, \citenamefont {Tung},
  \citenamefont {Rowley}, \citenamefont {Amaro}, \citenamefont {Chipot},
  \citenamefont {Wang},\ and\ \citenamefont {Gumbart}}]{Lee2016a}%
  \BibitemOpen
  \bibfield  {author} {\bibinfo {author} {\bibfnamefont {C.~T.}\ \bibnamefont
  {Lee}}, \bibinfo {author} {\bibfnamefont {J.}~\bibnamefont {Comer}}, \bibinfo
  {author} {\bibfnamefont {C.}~\bibnamefont {Herndon}}, \bibinfo {author}
  {\bibfnamefont {N.}~\bibnamefont {Leung}}, \bibinfo {author} {\bibfnamefont
  {A.}~\bibnamefont {Pavlova}}, \bibinfo {author} {\bibfnamefont {R.~V.}\
  \bibnamefont {Swift}}, \bibinfo {author} {\bibfnamefont {C.}~\bibnamefont
  {Tung}}, \bibinfo {author} {\bibfnamefont {C.~N.}\ \bibnamefont {Rowley}},
  \bibinfo {author} {\bibfnamefont {R.~E.}\ \bibnamefont {Amaro}}, \bibinfo
  {author} {\bibfnamefont {C.}~\bibnamefont {Chipot}}, \bibinfo {author}
  {\bibfnamefont {Y.}~\bibnamefont {Wang}}, \ and\ \bibinfo {author}
  {\bibfnamefont {J.~C.}\ \bibnamefont {Gumbart}},\ }\bibfield  {title}
  {\enquote {\bibinfo {title} {{Simulation-Based Approaches for Determining
  Membrane Permeability of Small Compounds}},}\ }\href {\doibase
  10.1021/acs.jcim.6b00022} {\bibfield  {journal} {\bibinfo  {journal} {Journal
  of Chemical Information and Modeling}\ }\textbf {\bibinfo {volume} {56}},\
  \bibinfo {pages} {721--733} (\bibinfo {year} {2016})}\BibitemShut {NoStop}%
\bibitem [{\citenamefont {Fathizadeh}\ and\ \citenamefont
  {Elber}(2019)}]{Fathizadeh2019}%
  \BibitemOpen
  \bibfield  {author} {\bibinfo {author} {\bibfnamefont {A.}~\bibnamefont
  {Fathizadeh}}\ and\ \bibinfo {author} {\bibfnamefont {R.}~\bibnamefont
  {Elber}},\ }\bibfield  {title} {\enquote {\bibinfo {title} {{Ion Permeation
  through a Phospholipid Membrane: Transition State, Path Splitting, and
  Calculation of Permeability}},}\ }\href {\doibase 10.1021/acs.jctc.8b00882}
  {\bibfield  {journal} {\bibinfo  {journal} {Journal of Chemical Theory and
  Computation}\ }\textbf {\bibinfo {volume} {15}},\ \bibinfo {pages} {720--730}
  (\bibinfo {year} {2019})}\BibitemShut {NoStop}%
\bibitem [{\citenamefont {Abraham}\ \emph {et~al.}(2015)\citenamefont
  {Abraham}, \citenamefont {Murtola}, \citenamefont {Schulz}, \citenamefont
  {P{\'{a}}ll}, \citenamefont {Smith}, \citenamefont {Hess},\ and\
  \citenamefont {Lindah}}]{Abraham2015}%
  \BibitemOpen
  \bibfield  {author} {\bibinfo {author} {\bibfnamefont {M.~J.}\ \bibnamefont
  {Abraham}}, \bibinfo {author} {\bibfnamefont {T.}~\bibnamefont {Murtola}},
  \bibinfo {author} {\bibfnamefont {R.}~\bibnamefont {Schulz}}, \bibinfo
  {author} {\bibfnamefont {S.}~\bibnamefont {P{\'{a}}ll}}, \bibinfo {author}
  {\bibfnamefont {J.~C.}\ \bibnamefont {Smith}}, \bibinfo {author}
  {\bibfnamefont {B.}~\bibnamefont {Hess}}, \ and\ \bibinfo {author}
  {\bibfnamefont {E.}~\bibnamefont {Lindah}},\ }\bibfield  {title} {\enquote
  {\bibinfo {title} {{Gromacs: High performance molecular simulations through
  multi-level parallelism from laptops to supercomputers}},}\ }\href {\doibase
  10.1016/j.softx.2015.06.001} {\bibfield  {journal} {\bibinfo  {journal}
  {SoftwareX}\ }\textbf {\bibinfo {volume} {1-2}},\ \bibinfo {pages} {19--25}
  (\bibinfo {year} {2015})},\ \Eprint {http://arxiv.org/abs/arXiv:1503.05249v1}
  {arXiv:arXiv:1503.05249v1} \BibitemShut {NoStop}%
\bibitem [{\citenamefont {P{\'{e}}rez-Hern{\'{a}}ndez}\ \emph
  {et~al.}(2013)\citenamefont {P{\'{e}}rez-Hern{\'{a}}ndez}, \citenamefont
  {Paul}, \citenamefont {Giorgino}, \citenamefont {{De Fabritiis}},\ and\
  \citenamefont {No{\'{e}}}}]{Perez-Hernandez2013}%
  \BibitemOpen
  \bibfield  {author} {\bibinfo {author} {\bibfnamefont {G.}~\bibnamefont
  {P{\'{e}}rez-Hern{\'{a}}ndez}}, \bibinfo {author} {\bibfnamefont
  {F.}~\bibnamefont {Paul}}, \bibinfo {author} {\bibfnamefont {T.}~\bibnamefont
  {Giorgino}}, \bibinfo {author} {\bibfnamefont {G.}~\bibnamefont {{De
  Fabritiis}}}, \ and\ \bibinfo {author} {\bibfnamefont {F.}~\bibnamefont
  {No{\'{e}}}},\ }\bibfield  {title} {\enquote {\bibinfo {title}
  {{Identification of slow molecular order parameters for Markov model
  construction}},}\ }\href {\doibase 10.1063/1.4811489} {\bibfield  {journal}
  {\bibinfo  {journal} {Journal of Chemical Physics}\ }\textbf {\bibinfo
  {volume} {139}} (\bibinfo {year} {2013}),\ 10.1063/1.4811489},\ \Eprint
  {http://arxiv.org/abs/arXiv:1302.6614v1} {arXiv:arXiv:1302.6614v1}
  \BibitemShut {NoStop}%
\bibitem [{\citenamefont {Husic}\ \emph {et~al.}(2016)\citenamefont {Husic},
  \citenamefont {McGibbon}, \citenamefont {Sultan},\ and\ \citenamefont
  {Pande}}]{Husic2016}%
  \BibitemOpen
  \bibfield  {author} {\bibinfo {author} {\bibfnamefont {B.~E.}\ \bibnamefont
  {Husic}}, \bibinfo {author} {\bibfnamefont {R.~T.}\ \bibnamefont {McGibbon}},
  \bibinfo {author} {\bibfnamefont {M.~M.}\ \bibnamefont {Sultan}}, \ and\
  \bibinfo {author} {\bibfnamefont {V.~S.}\ \bibnamefont {Pande}},\ }\bibfield
  {title} {\enquote {\bibinfo {title} {Optimized parameter selection reveals
  trends in markov state models for protein folding},}\ }\href {\doibase
  10.1063/1.4967809} {\bibfield  {journal} {\bibinfo  {journal} {The Journal of
  Chemical Physics}\ }\textbf {\bibinfo {volume} {145}},\ \bibinfo {pages}
  {194103} (\bibinfo {year} {2016})},\ \Eprint
  {http://arxiv.org/abs/https://doi.org/10.1063/1.4967809}
  {https://doi.org/10.1063/1.4967809} \BibitemShut {NoStop}%
\bibitem [{\citenamefont {Naritomi}\ and\ \citenamefont
  {Fuchigami}(2011)}]{Naritomi2011}%
  \BibitemOpen
  \bibfield  {author} {\bibinfo {author} {\bibfnamefont {Y.}~\bibnamefont
  {Naritomi}}\ and\ \bibinfo {author} {\bibfnamefont {S.}~\bibnamefont
  {Fuchigami}},\ }\bibfield  {title} {\enquote {\bibinfo {title} {{Slow
  dynamics in protein fluctuations revealed by time-structure based independent
  component analysis: The case of domain motions}},}\ }\href {\doibase
  10.1063/1.3554380} {\bibfield  {journal} {\bibinfo  {journal} {Journal of
  Chemical Physics}\ }\textbf {\bibinfo {volume} {134}} (\bibinfo {year}
  {2011}),\ 10.1063/1.3554380}\BibitemShut {NoStop}%
\bibitem [{\citenamefont {Beyerle}\ and\ \citenamefont
  {Guenza}(2021)}]{Beyerle2021c}%
  \BibitemOpen
  \bibfield  {author} {\bibinfo {author} {\bibfnamefont {E.~R.}\ \bibnamefont
  {Beyerle}}\ and\ \bibinfo {author} {\bibfnamefont {M.~G.}\ \bibnamefont
  {Guenza}},\ }\bibfield  {title} {\enquote {\bibinfo {title} {Identifying the
  leading dynamics of ubiquitin: a comparison between the tica and the le4pd
  slow fluctuations in amino acids? position},}\ }\href@noop {} {\bibfield
  {journal} {\bibinfo  {journal} {The Journal of Chemical Physics}\ }\textbf
  {\bibinfo {volume} {155}},\ \bibinfo {pages} {244108} (\bibinfo {year}
  {2021})}\BibitemShut {NoStop}%
\bibitem [{\citenamefont {E}\ and\ \citenamefont
  {Vanden-Eijnden}(2006)}]{E2006}%
  \BibitemOpen
  \bibfield  {author} {\bibinfo {author} {\bibfnamefont {W.}~\bibnamefont {E}}\
  and\ \bibinfo {author} {\bibfnamefont {E.}~\bibnamefont {Vanden-Eijnden}},\
  }\bibfield  {title} {\enquote {\bibinfo {title} {{Towards a theory of
  transition paths}},}\ }\href {\doibase 10.1007/s10955-005-9003-9} {\bibfield
  {journal} {\bibinfo  {journal} {Journal of Statistical Physics}\ }\textbf
  {\bibinfo {volume} {123}},\ \bibinfo {pages} {503--523} (\bibinfo {year}
  {2006})}\BibitemShut {NoStop}%
\bibitem [{\citenamefont {Noe}\ \emph {et~al.}(2009)\citenamefont {Noe},
  \citenamefont {Schutte}, \citenamefont {Vanden-Eijnden}, \citenamefont
  {Reich},\ and\ \citenamefont {Weikl}}]{Noe2009}%
  \BibitemOpen
  \bibfield  {author} {\bibinfo {author} {\bibfnamefont {F.}~\bibnamefont
  {Noe}}, \bibinfo {author} {\bibfnamefont {C.}~\bibnamefont {Schutte}},
  \bibinfo {author} {\bibfnamefont {E.}~\bibnamefont {Vanden-Eijnden}},
  \bibinfo {author} {\bibfnamefont {L.}~\bibnamefont {Reich}}, \ and\ \bibinfo
  {author} {\bibfnamefont {T.~R.}\ \bibnamefont {Weikl}},\ }\bibfield  {title}
  {\enquote {\bibinfo {title} {{Constructing the equilibrium ensemble of
  folding pathways from short off-equilibrium simulations}},}\ }\href {\doibase
  10.1073/pnas.0905466106} {\bibfield  {journal} {\bibinfo  {journal}
  {Proceedings of the National Academy of Sciences}\ }\textbf {\bibinfo
  {volume} {106}},\ \bibinfo {pages} {19011--19016} (\bibinfo {year}
  {2009})}\BibitemShut {NoStop}%
\bibitem [{\citenamefont {No{\'{e}}}\ \emph {et~al.}(2007)\citenamefont
  {No{\'{e}}}, \citenamefont {Horenko}, \citenamefont {Sch{\"{u}}tte},\ and\
  \citenamefont {Smith}}]{Noe2007}%
  \BibitemOpen
  \bibfield  {author} {\bibinfo {author} {\bibfnamefont {F.}~\bibnamefont
  {No{\'{e}}}}, \bibinfo {author} {\bibfnamefont {I.}~\bibnamefont {Horenko}},
  \bibinfo {author} {\bibfnamefont {C.}~\bibnamefont {Sch{\"{u}}tte}}, \ and\
  \bibinfo {author} {\bibfnamefont {J.~C.}\ \bibnamefont {Smith}},\ }\bibfield
  {title} {\enquote {\bibinfo {title} {{Hierarchical analysis of conformational
  dynamics in biomolecules: Transition networks of metastable states}},}\
  }\href {\doibase 10.1063/1.2714539} {\bibfield  {journal} {\bibinfo
  {journal} {Journal of Chemical Physics}\ }\textbf {\bibinfo {volume} {126}},\
  \bibinfo {pages} {155102} (\bibinfo {year} {2007})}\BibitemShut {NoStop}%
\bibitem [{\citenamefont {Prinz}\ \emph {et~al.}(2011)\citenamefont {Prinz},
  \citenamefont {Wu}, \citenamefont {Sarich}, \citenamefont {Keller},
  \citenamefont {Senne}, \citenamefont {Held}, \citenamefont {Chodera},
  \citenamefont {Sch{\"{u}}tte},\ and\ \citenamefont {No{\'{e}}}}]{Prinz2011}%
  \BibitemOpen
  \bibfield  {author} {\bibinfo {author} {\bibfnamefont {J.-H.}\ \bibnamefont
  {Prinz}}, \bibinfo {author} {\bibfnamefont {H.}~\bibnamefont {Wu}}, \bibinfo
  {author} {\bibfnamefont {M.}~\bibnamefont {Sarich}}, \bibinfo {author}
  {\bibfnamefont {B.}~\bibnamefont {Keller}}, \bibinfo {author} {\bibfnamefont
  {M.}~\bibnamefont {Senne}}, \bibinfo {author} {\bibfnamefont
  {M.}~\bibnamefont {Held}}, \bibinfo {author} {\bibfnamefont {J.~D.}\
  \bibnamefont {Chodera}}, \bibinfo {author} {\bibfnamefont {C.}~\bibnamefont
  {Sch{\"{u}}tte}}, \ and\ \bibinfo {author} {\bibfnamefont {F.}~\bibnamefont
  {No{\'{e}}}},\ }\bibfield  {title} {\enquote {\bibinfo {title} {{Markov
  models of molecular kinetics: Generation and validation}},}\ }\href {\doibase
  10.1063/1.3565032} {\bibfield  {journal} {\bibinfo  {journal} {The Journal of
  Chemical Physics}\ }\textbf {\bibinfo {volume} {134}},\ \bibinfo {pages}
  {174105} (\bibinfo {year} {2011})}\BibitemShut {NoStop}%
\bibitem [{\citenamefont {Scherer}\ \emph {et~al.}(2015)\citenamefont
  {Scherer}, \citenamefont {Trendelkamp-Schroer}, \citenamefont {Paul},
  \citenamefont {P{\'{e}}rez-Hern{\'{a}}ndez}, \citenamefont {Hoffmann},
  \citenamefont {Plattner}, \citenamefont {Wehmeyer}, \citenamefont {Prinz},\
  and\ \citenamefont {No{\'{e}}}}]{Scherer2015}%
  \BibitemOpen
  \bibfield  {author} {\bibinfo {author} {\bibfnamefont {M.~K.}\ \bibnamefont
  {Scherer}}, \bibinfo {author} {\bibfnamefont {B.}~\bibnamefont
  {Trendelkamp-Schroer}}, \bibinfo {author} {\bibfnamefont {F.}~\bibnamefont
  {Paul}}, \bibinfo {author} {\bibfnamefont {G.}~\bibnamefont
  {P{\'{e}}rez-Hern{\'{a}}ndez}}, \bibinfo {author} {\bibfnamefont
  {M.}~\bibnamefont {Hoffmann}}, \bibinfo {author} {\bibfnamefont
  {N.}~\bibnamefont {Plattner}}, \bibinfo {author} {\bibfnamefont
  {C.}~\bibnamefont {Wehmeyer}}, \bibinfo {author} {\bibfnamefont {J.~H.}\
  \bibnamefont {Prinz}}, \ and\ \bibinfo {author} {\bibfnamefont
  {F.}~\bibnamefont {No{\'{e}}}},\ }\bibfield  {title} {\enquote {\bibinfo
  {title} {{PyEMMA 2: A Software Package for Estimation, Validation, and
  Analysis of Markov Models}},}\ }\href {\doibase 10.1021/acs.jctc.5b00743}
  {\bibfield  {journal} {\bibinfo  {journal} {Journal of Chemical Theory and
  Computation}\ }\textbf {\bibinfo {volume} {11}},\ \bibinfo {pages}
  {5525--5542} (\bibinfo {year} {2015})}\BibitemShut {NoStop}%
\bibitem [{\citenamefont {Berezhkovskii}\ and\ \citenamefont
  {Szabo}(2004)}]{Berezhkovskii2004}%
  \BibitemOpen
  \bibfield  {author} {\bibinfo {author} {\bibfnamefont {A.}~\bibnamefont
  {Berezhkovskii}}\ and\ \bibinfo {author} {\bibfnamefont {A.}~\bibnamefont
  {Szabo}},\ }\bibfield  {title} {\enquote {\bibinfo {title} {Ensemble of
  transition states for two-state protein folding from the eigenvectors of rate
  matrices},}\ }\href {\doibase 10.1063/1.1802674} {\bibfield  {journal}
  {\bibinfo  {journal} {The Journal of Chemical Physics}\ }\textbf {\bibinfo
  {volume} {121}},\ \bibinfo {pages} {9186--9187} (\bibinfo {year} {2004})},\
  \Eprint {http://arxiv.org/abs/https://doi.org/10.1063/1.1802674}
  {https://doi.org/10.1063/1.1802674} \BibitemShut {NoStop}%
\bibitem [{\citenamefont {Buchete}\ and\ \citenamefont
  {Hummer}(2008)}]{Buchete2008}%
  \BibitemOpen
  \bibfield  {author} {\bibinfo {author} {\bibfnamefont {N.~V.}\ \bibnamefont
  {Buchete}}\ and\ \bibinfo {author} {\bibfnamefont {G.}~\bibnamefont
  {Hummer}},\ }\bibfield  {title} {\enquote {\bibinfo {title} {{Coarse master
  equations for peptide folding dynamics}},}\ }\href {\doibase
  10.1021/jp0761665} {\bibfield  {journal} {\bibinfo  {journal} {Journal of
  Physical Chemistry B}\ }\textbf {\bibinfo {volume} {112}},\ \bibinfo {pages}
  {6057--6069} (\bibinfo {year} {2008})}\BibitemShut {NoStop}%
\bibitem [{\citenamefont {Cardenas}\ and\ \citenamefont
  {Elber}(2013)}]{Cardenas2013}%
  \BibitemOpen
  \bibfield  {author} {\bibinfo {author} {\bibfnamefont {A.~E.}\ \bibnamefont
  {Cardenas}}\ and\ \bibinfo {author} {\bibfnamefont {R.}~\bibnamefont
  {Elber}},\ }\bibfield  {title} {\enquote {\bibinfo {title} {Computational
  study of peptide permeation through membrane: searching for hidden slow
  variables},}\ }\href {\doibase 10.1080/00268976.2013.842010} {\bibfield
  {journal} {\bibinfo  {journal} {Molecular Physics}\ }\textbf {\bibinfo
  {volume} {111}},\ \bibinfo {pages} {3565--3578} (\bibinfo {year} {2013})},\
  \bibinfo {note} {pMID: 26203198},\ \Eprint
  {http://arxiv.org/abs/https://doi.org/10.1080/00268976.2013.842010}
  {https://doi.org/10.1080/00268976.2013.842010} \BibitemShut {NoStop}%
\bibitem [{\citenamefont {Bussi}\ and\ \citenamefont
  {Tribello}(2019)}]{Bussi2019}%
  \BibitemOpen
  \bibfield  {author} {\bibinfo {author} {\bibfnamefont {G.}~\bibnamefont
  {Bussi}}\ and\ \bibinfo {author} {\bibfnamefont {G.~A.}\ \bibnamefont
  {Tribello}},\ }\bibfield  {title} {\enquote {\bibinfo {title} {Analyzing and
  biasing simulations with plumed},}\ }in\ \href@noop {} {\emph {\bibinfo
  {booktitle} {Biomolecular Simulations}}}\ (\bibinfo  {publisher} {Springer},\
  \bibinfo {year} {2019})\ pp.\ \bibinfo {pages} {529--578}\BibitemShut
  {NoStop}%
\bibitem [{\citenamefont {Sittel}\ and\ \citenamefont
  {Stock}(2018)}]{Sittel2018}%
  \BibitemOpen
  \bibfield  {author} {\bibinfo {author} {\bibfnamefont {F.}~\bibnamefont
  {Sittel}}\ and\ \bibinfo {author} {\bibfnamefont {G.}~\bibnamefont {Stock}},\
  }\bibfield  {title} {\enquote {\bibinfo {title} {Perspective: Identification
  of collective variables and metastable states of protein dynamics},}\ }\href
  {\doibase 10.1063/1.5049637} {\bibfield  {journal} {\bibinfo  {journal} {The
  Journal of Chemical Physics}\ }\textbf {\bibinfo {volume} {149}},\ \bibinfo
  {pages} {150901} (\bibinfo {year} {2018})},\ \Eprint
  {http://arxiv.org/abs/https://doi.org/10.1063/1.5049637}
  {https://doi.org/10.1063/1.5049637} \BibitemShut {NoStop}%
\bibitem [{\citenamefont {Allen}, \citenamefont {Valeriani},\ and\
  \citenamefont {Ten~Wolde}(2009)}]{Allen2009}%
  \BibitemOpen
  \bibfield  {author} {\bibinfo {author} {\bibfnamefont {R.~J.}\ \bibnamefont
  {Allen}}, \bibinfo {author} {\bibfnamefont {C.}~\bibnamefont {Valeriani}}, \
  and\ \bibinfo {author} {\bibfnamefont {P.~R.}\ \bibnamefont {Ten~Wolde}},\
  }\bibfield  {title} {\enquote {\bibinfo {title} {Forward flux sampling for
  rare event simulations},}\ }\href@noop {} {\bibfield  {journal} {\bibinfo
  {journal} {Journal of physics: Condensed matter}\ }\textbf {\bibinfo {volume}
  {21}},\ \bibinfo {pages} {463102} (\bibinfo {year} {2009})}\BibitemShut
  {NoStop}%
\bibitem [{\citenamefont {{Towns}}\ \emph {et~al.}(2014)\citenamefont
  {{Towns}}, \citenamefont {{Cockerill}}, \citenamefont {{Dahan}},
  \citenamefont {{Foster}}, \citenamefont {{Gaither}}, \citenamefont
  {{Grimshaw}}, \citenamefont {{Hazlewood}}, \citenamefont {{Lathrop}},
  \citenamefont {{Lifka}}, \citenamefont {{Peterson}}, \citenamefont
  {{Roskies}}, \citenamefont {{Scott}},\ and\ \citenamefont
  {{Wilkins-Diehr}}}]{Towns2014}%
  \BibitemOpen
  \bibfield  {author} {\bibinfo {author} {\bibfnamefont {J.}~\bibnamefont
  {{Towns}}}, \bibinfo {author} {\bibfnamefont {T.}~\bibnamefont
  {{Cockerill}}}, \bibinfo {author} {\bibfnamefont {M.}~\bibnamefont
  {{Dahan}}}, \bibinfo {author} {\bibfnamefont {I.}~\bibnamefont {{Foster}}},
  \bibinfo {author} {\bibfnamefont {K.}~\bibnamefont {{Gaither}}}, \bibinfo
  {author} {\bibfnamefont {A.}~\bibnamefont {{Grimshaw}}}, \bibinfo {author}
  {\bibfnamefont {V.}~\bibnamefont {{Hazlewood}}}, \bibinfo {author}
  {\bibfnamefont {S.}~\bibnamefont {{Lathrop}}}, \bibinfo {author}
  {\bibfnamefont {D.}~\bibnamefont {{Lifka}}}, \bibinfo {author} {\bibfnamefont
  {G.~D.}\ \bibnamefont {{Peterson}}}, \bibinfo {author} {\bibfnamefont
  {R.}~\bibnamefont {{Roskies}}}, \bibinfo {author} {\bibfnamefont {J.~R.}\
  \bibnamefont {{Scott}}}, \ and\ \bibinfo {author} {\bibfnamefont
  {N.}~\bibnamefont {{Wilkins-Diehr}}},\ }\bibfield  {title} {\enquote
  {\bibinfo {title} {Xsede: Accelerating scientific discovery},}\ }\href@noop
  {} {\bibfield  {journal} {\bibinfo  {journal} {Computing in Science
  Engineering}\ }\textbf {\bibinfo {volume} {16}},\ \bibinfo {pages} {62--74}
  (\bibinfo {year} {2014})}\BibitemShut {NoStop}%
\end{thebibliography}

%

\newpage
\begin{figure*}[p]
    \centering
    \includegraphics[width=0.7\textwidth]{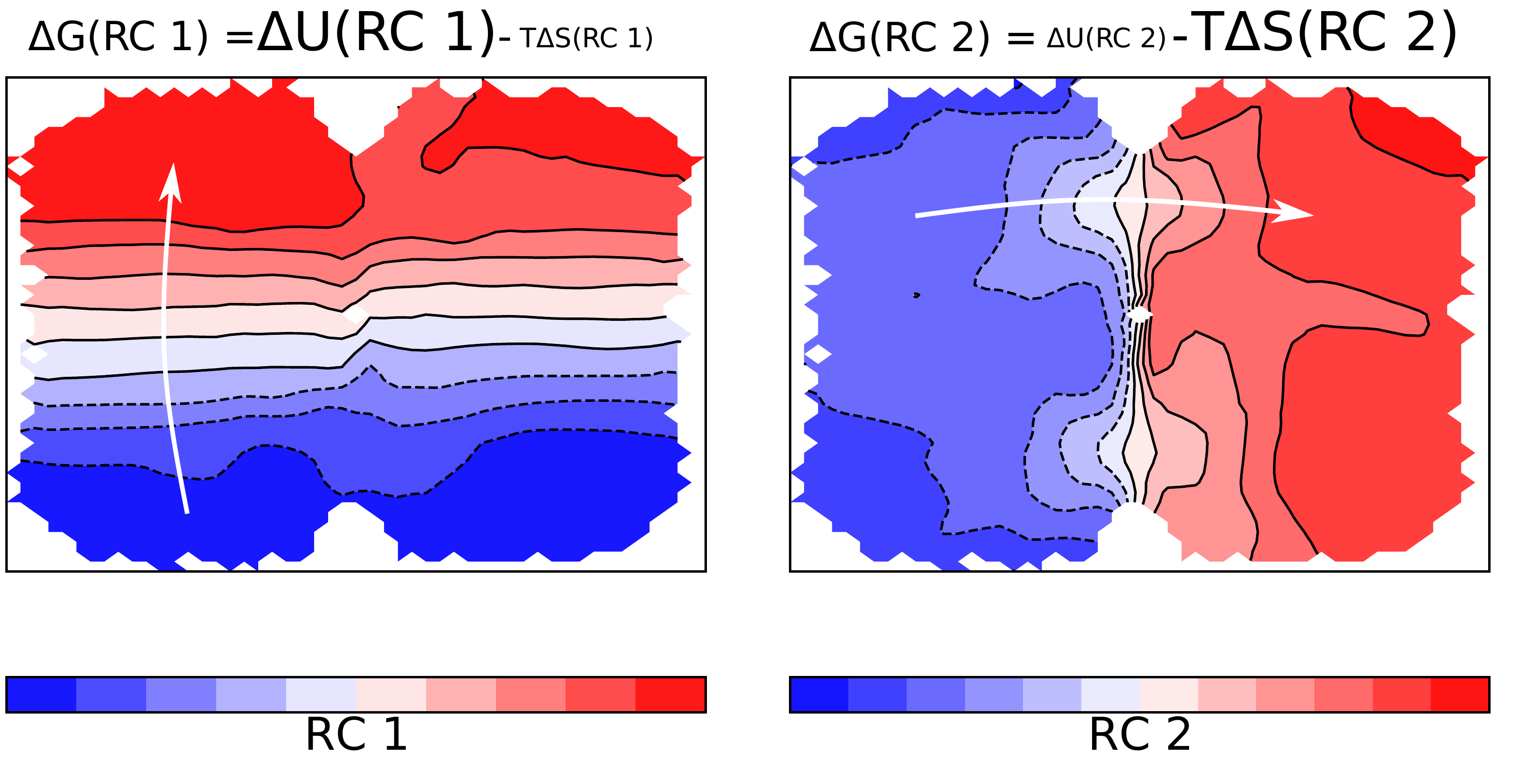}
    \captionsetup{labelformat=empty}
    \caption{For Table of Contents Only}
\end{figure*}
\end{document}